\documentclass[12pt,preprint]{aastex}
\usepackage{amsmath}
\usepackage{color}
\definecolor{grey}{rgb}{0.5,0.5,0.5}
\definecolor{red}{rgb}{0.8,0.0,0.0}
\definecolor{blue}{rgb}{0.0,0.0,0.8}

\newcommand{\Vp}{V_p}
\newcommand{\Vsys}{V_\mathrm{sys}}
\newcommand{\Texp}{T_\mathrm{exp}}
\newcommand{\Vrot}{V_\mathrm{rot}}

\newcommand{\Thput}{\eta}
\newcommand{\fsky}{f_\mathrm{sky}}
\newcommand{\Ssky}{S_\mathrm{sky}}
\newcommand{\Isky}{I_\mathrm{sky}}

\newcommand{\Cinst}{C_\mathrm{raw}}
\newcommand{\Csp}{C_\mathrm{ps}}
\newcommand{\NCCF}{N_\mathrm{eff}}
\newcommand{\np}{n_\mathrm{p}}
\newcommand{\speckle}{\mathrm{speckle}}
\newcommand{\tot}{\mathrm{tot}}
\newcommand{\Fobs}{\mathcal{F}}
\newcommand{\Fpest}{\Fobs_{p}}
\newcommand{\Fpesti}{\Fobs_{p}}
\newcommand{\ccp}{g_t}
\newcommand{\ulim}{u_\mathrm{lim}}

\slugcomment{}
\shortauthors{Kawahara and Hirano}
\shorttitle{Doppler-shifted Water and Oxygen of ELPs}
\begin{document}
\title{Characterizing Earth-like Planets  using A Combination of High-Dispersion Spectroscopy and High-Contrast Instruments: Doppler-shifted Water and Oxygen Lines}

\author{Hajime Kawahara\altaffilmark{1} and Teruyuki Hirano\altaffilmark{2}}
\altaffiltext{1}{Department of Earth and Planetary Science, The University of Tokyo, 
Tokyo 113-0033, Japan}
\altaffiltext{2}{Department of Earth and Planetary Sciences, Tokyo Institute of Technology, Tokyo 152-8550, Japan}
\email{kawahara@eps.s.u-tokyo.ac.jp}

\begin{abstract}
Future radial velocity, astrometric, and direct-imaging surveys will find nearby Earth-sized planets within the habitable zone in the near future. How can we search for water and oxygen in those nontransiting planets? We show that a combination of high-dispersion spectroscopic and coronagraphic techniques is a promising technique to detect molecular lines imprinted in the scattered light of Earth-like planets (ELPs). In this method, the planetary signals are spectroscopically separated from telluric absorption by using the Doppler shift. Assuming a long observing campaign (a 10-day exposure) using a high-dispersion spectrometer (R=50,000) with speckle suppression on a 30-m telescope, we simulate the spectra from ELPs around M dwarfs (whose stellar effective temperature is 2750-3750 K) at 5 pc. Performing a cross-correlation analysis with the spectral template of the molecular lines, we find that raw contrasts of $10^{-4}$ and $10^{-5}$ (using Y, J, and H bands) are required to detect water vapor at the 3 $\sigma$ and 16 $\sigma$ levels, respectively, for $T_\star$=3000 K. The raw contrast of $10^{-5}$ is required for a 6 $\sigma$ detection of the oxygen 1.27 $\mu$m band. We also examine possible systematics, incomplete speckle subtraction, and the correction for telluric lines. When those are not perfect, a telluric water signal appears in the cross-correlation function. However, we find the planetary signal is separated from that resulting from the velocity difference. We also find that the intrinsic water lines in the Phoenix spectra are too weak to affect the results for water detection. This method does not require any additional post-processing and is less sensitive to telluric noise than low-resolution spectroscopy. We conclude that a combination of high-dispersion spectroscopy and high-contrast instruments can be a powerful means to characterize ELPs in the extremely large telescope era.
\end{abstract}

\section{Introduction}
 Spectroscopic detection of atmospheric water vapor,oxygen,and other biosignatures will be the first step to search for exolife on exoplanets.Direct imaging from space has been regarded as a promising way to find Earth-like planets (ELPs) in the habitable zone (HZ).In this case,low-resolution spectroscopy (typically of resolving power $R \sim 100 - 1000$) will be a means of searching for water vapor and other biosignatures \citep[e.g.][]{2002AsBio...2..153D,2006ApJ...644..551T,2007ApJ...658..598K}. Strong molecular features including numerous water bands,the oxygen 0.76$\mu$m band,and the ozone 10 $\mu$m band are the targets of biosignature searches for the proposed dedicated space missions.

Low-resolution spectroscopy of direct imaging with extremely large telescopes (ELTs) has also been considered in the context of the search for oxygen in ELPs around late-type stars \citep{2012ApJ...758...13K} because the contrast of habitable planets around a M-type star is a factor of 100 greater than that around G-type stars \citep[e.g.][]{2010SPIE.7735E..84M,2012ApJ...758...13K,2012SPIE.8447E..1XG,2013A&A...551A..99C,2014arXiv1407.5099M}. However, sophisticated post-processing is required to reach the typical planet-star contrast of $\sim 10^{-8}$ from the raw contrast of $10^{-4} - 10^{-5}$ for Earth-sized planets in the HZ. Accurate calibrations of the time-variable nighttime airglow spectrum and the transmittance of our Earth are also crucial for low-resolution spectroscopy \citep{2012ApJ...758...13K}. Furthermore, molecular detection with low-resolution spectroscopy requires full spectrum modeling of the planets because many thin lines are blended in the spectrum.

In principle, high-dispersion spectroscopy can solve those problems. Simultaneous identification of multiple rotational-vibrational molecular lines makes detection of the molecule robust. In our solar system, \citet{1963ApJ...137.1319S} first reported the detection of Martian water vapor using the Mount Wilson 100-inch reflector in the near-infrared (NIR) band. Performing a high-dispersion observation, they could separate the Martian water lines with a Doppler shift $\Delta V=15$ km/s from the telluric lines. To date, Doppler-shifted water lines in hot Jupiters have been detected despite the weak signal with the planet-star contrast being $>10^{-3}$ \citep{2013MNRAS.436L..35B,2014ApJ...783L..29L,2014A&A...565A.124B}. \citet{2013ApJ...764..182S} proposed using high-dispersion transmission spectroscopy to detect the oxygen 0.76$\mu m$ lines for transiting ELPs around nearby late-type stars. \citet{2014ApJ...781...54R} further studied its feasibility and concluded that the detection will be feasible with ELTs for an Earth analog within 8 pc, in the most optimistic cases. As demonstrated by them, high-dispersion transmission spectroscopy is a promising means to characterize transiting ELPs.

 Recent progress in radial velocity (RV) surveys has significantly improved their precision and a number of RV instruments have been planned for nearby planet surveys \citep[ e.g. CARMENES, CRIRES, ESPRESSO, HPF, IRD, SPIRou; ][]{2010SPIE.7735E..13Q, 2010SPIE.7735E..0FP,kotani2014infrared,2014arXiv1406.6992A}. An RV precision of $\sim 1$m/s in the NIR covers an Earth-sized planet around nearby late-type stars. The unprecedented required precision of 10 cm/s in the visible band of ESPRESSO will reach rocky planets around solar-type stars \citep{2010SPIE.7735E..0FP}. After the discovery of those planets in the near future, the question arises at to how we can characterize nearby habitable planet candidates to search for exolife. It is important to explore the characterization methods applicable to nontransiting ELPs. The first detection of Doppler-shifted molecular lines for a nontransiting planet was achieved by \citet{2014Natur.509...63S}. They performed high-dispersion slit spectroscopy of the self-luminous direct-imaged planet, beta Pictorias b, and detected Doppler-shifted carbon monoxide lines at the position of the planet. Using adaptive optics, they could obtain contrast at the planet that was a factor of 9-30 greater than that in the integrated light. Utilizing a combination of a high-contrast instrument and high-dispersion spectroscopy will be one of the promising ways for characterizing exoplanets \citep[see also][]{2014ApJS..212...27K, 2014arXiv1409.3087B}.

In this paper, we consider high-dispersion spectroscopic detection of water vapor and oxygen in ELPs (Earth analog) assuming future high-contrast and high-dispersion instruments on ELTs. In particular, we focus on ELPs around late-type stars ($T_\star \sim 3000$ K) because these targets are expected to have better contrast ($\sim 10^{-8}$) than that around solar-type stars. There are several important differences between the current high-dispersion detections for hot Jupiters and the method for ELPs we consider. First, the detection method is primarily direct imaging: The detection limit strongly depends on the performance of the high-contrast instruments, so we examine the required performance of the high-contrast instruments for molecular detection. Second, the light from the ELP we consider is scattered light, not emission light. This fact implies that stellar lines can contaminate the planet spectrum. Intrinsic water lines in the stellar spectra of late-type stars can cause false positives, while stellar intrinsic lines in the scattered lines can also be regarded as the signal of the planet, as pointed by \cite{2013MNRAS.436.1215M}. Third, telluric contamination is different from the case of hot Jupiters because the ELT signal + speckle system is much fainter than the hot Jupiter + star system. The nighttime airglow effect should be considered. Moreover, the suitable band to observe is different: The NIR band will be assumed for water and oxygen detection because it is the most suitable for wavefront correction by the extreme AO (ExAO), whereas the current water detection of hot Jupiters uses a longer wavelength (L band). To investigate the feasibility of the method including these effects, we perform detailed simulations using the radiative transfer of the Earth.

 The paper is organized as follows. In Section 2, we provide the definitions of the Doppler-shifted water, oxygen, and scattered stellar lines and describe the statistics for a given planetary system and instrumental and observational conditions. In Section 3, the spectra at the planet position are created by including the planetary scattered spectrum simulated by the radiative transfer code, the stellar speckle, the nighttime airglow, and the transmittance of our Earth. A cross-correlation analysis of the mock spectra and the spectral templates of the molecular lines is performed to study the feasibility. In Section 4, we show the results of the cross-correlation analysis for water, oxygen, carbon dioxide, and stellar intrinsic lines scattered by a planet. We also investigate the sensitivity of the results to the systematics. In Section 5, we summarize our results.

\section{High-Dispersion Observation of Earth-like Planets with Ground-Based Telescopes}

 In this section, we explain what type of observation we assume and show how this method works using orders-of-magnitude arguments. We consider high-dispersion spectroscopy of ELPs after speckle suppression. The planetary signals are spectroscopically separated from the stellar speckle and the telluric lines. The spectral range is restricted to Y, J, and H bands because of the balance of the inner working angle and ease of wavefront control. We assume an observing campaign of one planet with a long exposure time (~10 days) because the RV or other planet surveys will have detected the target planet in the HZ in the near future.

We also assume that the aperture extracted for the spectrum analysis is fixed at the planet position. If the planet can be detected by direct imaging with post-processing, as a result, the position angle of the planet is known, and one can perform slit spectroscopy or fiber spectroscopy at the planet position. If we do not utilize direct-imaging detection, we require integral field spectroscopy with high spectral resolution because we only know the angular separation of the planet and the host star in this case. However, we will not consider the detector type any further and concentrate on the signals at the planetary position.

\subsection{Classification of the Molecular Lines}

 To separate the planetary signals, we utilize the relative velocities of the star, the planet, and the observer frame. The origins of the molecular lines in the integrated spectrum are characterized by the relative velocity as shown in Table \ref{tab:lines}. Doppler-shifted molecular lines originating from the planetary atmosphere (planetary lines) have a relative velocity that is the sum of the velocity owing to the orbital motion of the planet, $\Vp$ and the peculiar velocity of the system, $\Vsys$, with respect to the observer (the relative velocity between the star and the Sun and the orbital and spin motion of the Earth). The lines originating from the stellar spectrum are imprinted in the integrated spectrum in two ways with different Doppler velocities. One is imprinted in the planetary spectrum because the scattered light of the planet is originally star light. Those lines are regarded as the planetary signals as well as the planetary lines with the peculiar velocity $v = \Vsys + \Vp$. We refer to those lines as scattered stellar lines. The other is simply from the speckle, i.e., contamination of the central starlight (speckle stellar lines), which has the peculiar velocity of the system, $v=\Vsys$. Hence this velocity difference between the direct star light and the scattered star light is . Scattered stellar lines have been proposed as a tracer of scattered light from exoplanets by \citet{2013MNRAS.436.1215M}. Finally, telluric water vapor, oxygen, and other species contribute numerous absorption lines to the spectrum and OH and $\mathrm{O}_2$ emit peculiar lines, i.e., the nighttime airglow. Those lines have the velocity of the observer frame, i.e., $v=0$ in our definition. We refer to those lines as telluric lines.

The radial velocity resulting from planetary orbital motion is expressed as
\begin{eqnarray}
  \Vp =  v_\mathrm{orb} \sin{i}  \, \sin (\omega t),
\end{eqnarray}
for a circular orbit, where $v_\mathrm{orb}$ indicates the orbital velocity of the planet, $i$ is the orbital inclination, and $\omega$ is the orbital angular velocity. The typical orbital velocity at the HZ \citep{2013ApJ...765..131K} around main sequence stars is 20-50 km/s for stellar masses of $M_\star = 0.1-1 \, M_\odot$. The orbital velocity at the HZ generally increases as the stellar mass decreases. Hence, high resolution spectroscopy with $R \sim 10^5$ can distinguish the velocity difference $\Vp$ unless the system is almost face-on ($i \lesssim 20-30^\circ$).

\begin{table}[!htb]
\begin{center}
\caption{Classification of molecular lines in the integrated light \label{tab:lines}}
  \begin{tabular}{ccc}
   \hline\hline
lines & velocity & origin \\
\hline
 Planetary lines & $\Vp+\Vsys$ &  $A(\lambda)$ in  $f_\mathrm{p} (\lambda)$\\
 Scattered stellar lines & $\Vp+\Vsys$ & $F_\star(\lambda)$ in $f_\mathrm{p} (\lambda)$ \\
 Speckle stellar lines & $\Vsys$ & $f_\speckle (\lambda)$\\
 Telluric lines  & $0$ & $T(\lambda)$ \\
 Airglow & $0$ & $\fsky(\lambda)$\\
\end{tabular}
\end{center}
\end{table}

The spectrum at the planet position is modeled as 
\begin{eqnarray}
\label{eq:totalflux}
f_\tot(\lambda) &=& T(\lambda) f_\mathrm{p}(\lambda) + f_\speckle (\lambda) + \fsky (\lambda) + f_n(\lambda),
\end{eqnarray}
where $T(\lambda)$ is the transmittance of our Earth, $f_\mathrm{p}(\lambda)$ and $f_\speckle (\lambda)$ indicate the Doppler-shifted planetary and speckle stellar spectra. The nighttime airglow and instrumental noise, such as readout noise, are denoted by $\fsky (\lambda)$ and $f_n (\lambda)$, respectively. Let us explain each term in detail below.

The planetary spectrum is shifted as 
\begin{eqnarray}
\label{eq:planet}
f_\mathrm{p}(\lambda) \Delta \lambda &\equiv& F_\mathrm{p} (\lambda \delta) \frac{\Delta \lambda }{\delta}, \nonumber \\
\delta &\equiv& \left(1 + \frac{\Vp+\Vsys}{c} \right),
\end{eqnarray}
 where $F_\mathrm{p} (\lambda)$ is the rest-frame planetary spectrum. One of the dominant noises is speckle noise from stellar light, in other words, contamination from stellar light at the planet position on the detector. This leakage from the stellar light is expressed as
\begin{eqnarray}
\label{eq:speckle}
f_\speckle(\lambda) \Delta \lambda &\equiv& \Cinst (\lambda) \, T(\lambda) \, F_\star(\lambda \delta^{\prime}) \frac{\Delta \lambda}{\delta^\prime}, \nonumber \\
\delta^{\prime} &\equiv& \left(1 +\frac{\Vsys}{c} \right).
\end{eqnarray}
 where $\Cinst (\lambda)$ is the raw contrast with the point spread function (PSF) circle at the planet position\footnote{ Strictly, the incoming stellar spectrum to the planet slightly differs from the stellar spectrum we observe (i.e., it affects the shape of the speckles) because the relative rotation velocity between the stellar spin and the orbital motion of the planet broadens the scattered spectrum. We ignore this difference and regard them as the same spectra in our simulation because the relative rotation velocity is of the same order of the stellar rotation or smaller for our case. We note that one cannot ignore this effect for very closely orbiting planets whose periods are several days. In this paper, we also do not consider the slight shift of the planet radial velocity resulting from spin rotation, which can be used to determine the planetary obliquity \cite{kawahara12}.} and $F_\star(\lambda)$ is the rest-frame stellar spectrum. In this paper, we assume a constant value of the raw contrast: $\Cinst (\lambda) = \Cinst$.

\subsection{Estimating the Required Flux and the Speckle Suppression to Detect Molecular Lines}
The total luminosity of the scattered light of an Earth-sized planet in the HZ does not depend much on the stellar type because the incoming energy required to maintain habitability is approximately constant and the scattered light is part of its incoming energy. An Earth analog with $R_\oplus$ and A=0.3 has magnides of 26–27 in the NIR (Y, J, and H) bands at a distance of  $d=$ 5 pc.

The number of photons in $\Delta \lambda = \lambda/R$ (the width of the spectrum bins) for $R=10^5$ is expressed as  
\begin{eqnarray}
\np \Delta \lambda &\approx& 500 \left(\frac{F}{F_{-19}}\right) \left(\frac{\Thput}{0.1}\right) \left(\frac{\Texp}{10 \,\mathrm{days}}\right) \left(\frac{D}{30 \,\mathrm{m}}\right)^2 \left(\frac{R}{10^5}\right)^{-1}  \nonumber \\
&\,& \mathrm{[cts/bin]},
\end{eqnarray}
where $n_p$ is the planetary flux in photon count, $F_\mathrm{-19} = 10^{-19} \mathrm{[erg/s/cm^2/nm]}$, corresponding to J-band magnitude of 26.3 and $\Thput$ is the instrumental total throughput, $D$ is the telescope diameter, $\Texp$ is the exposure time and $R$ is a spectral resolving power. We adopted $\Thput=$ 10 \% for the high-dispersion instrument and a 10-day exposure time. 

The extent to which each molecular line contributes to the signal depends on the line depth, contamination of neighbor lines, and the transmittance of our Earth. Here, we define the effective number of available molecular lines, $\NCCF$, to include those effects. Then, we approximate the signal-to-noise ratio of the detection of planetary molecules as
\begin{eqnarray}
\label{eq:snest}
\mathrm{(S/N)}_\mathrm{det} \approx r \sqrt{\NCCF} \mathrm{(S/N)_{spectra}},
\end{eqnarray}
where $r$ is the ratio of the planetary flux and the total flux in the aperture,  
\begin{eqnarray}
\label{eq:snestr}
r = \frac{\np}{\Cinst n_\star + \np + \Ssky \Delta \Omega },
\end{eqnarray}
where $n_\star$ is the stellar flux in photon counts, $ \Cinst n_\star$ is the photon count of the speckle noise within the aperture $\Delta \Omega$ and $\Ssky$ is the sky surface brightness from the nighttime airglow. The raw contrast of the high-contrast instruments is denoted by $\Cinst$, which is defined by the ratio of the total photon count of an on-axis source and the photon count within the PSF circle at the planet position. We assume that the aperture has the same size as the PSF. In this case, the definition of $\Cinst$ provides a similar value of the PSF (raw) contrast defined by \citet{2005ApJ...629..592G}\footnote{In \citet{2005ApJ...629..592G}, the PSF contrast is defined as  the ratio of the intensity at the point of the target , $I({\bf \theta})$ to the intensity at the PSF center, $I(0)$, i.e. $C_\mathrm{PSF} \equiv I({\bf \theta})/I(0)$. Using this notation, our definition of $\Cinst$ can be written as $\Cinst = \int_{\Delta \Omega} \, d {\bf \theta^\prime} \, I({\bf \theta^\prime - \theta} )/\int \, d {\bf \theta^\prime} \, I({\bf \theta^\prime})$. By assuming that $\Delta \Omega$ is the PSF size, $ \int_{\Delta \Omega} \, d {\bf \theta^\prime} \, I({\bf \theta^\prime - \theta} ) \approx I({\bf \theta}) \Delta \Omega$ and $\int \, d {\bf \theta^\prime} \, I({\bf \theta^\prime}) \approx I(0) \Delta \Omega$. Then, we obtain $\Cinst \approx C_\mathrm{PSF}$. }. The raw contrast $\Cinst$ is generally worse than the final contrast for the direct imaging after PSF subtractions such as Angular Differential Imaging, Simultaneous Spectroscopic Differential Imaging, and Locally Optimized Combination of Images.

If the noise is dominated by the speckle, equation (\ref{eq:snest}) yields
\begin{eqnarray}
\label{eq:snestaa}
\mathrm{(S/N)}_\mathrm{det} &\approx& \frac{\Csp}{\Cinst} \sqrt{\NCCF} \mathrm{(S/N)_{spectra}} \nonumber  \\
&=& \sqrt{\frac{\Csp}{\Cinst} \NCCF \np \Delta \lambda}, 
\end{eqnarray}
where  $\Csp \equiv F_\mathrm{p}/F_\star$ is the planet-star contrast. 

For the scattered light, the planetary flux is expressed as 
\begin{eqnarray}
\label{eq:reffp}
F_\mathrm{p} (\lambda) = \frac{2}{3} \phi(\beta) \left(\frac{R_p}{a}\right)^2 A(\lambda)  F_\star(\lambda), 
\end{eqnarray}
where $R_p$ is the planet radius, $\phi (\beta)$ is the phase function as a function of the observer-star-planet angle $\beta$, and $A(\lambda)$ is the spherical albedo \citep{1975lpsa.book.....S}. Assuming isotropic scattering, we can obtain $\phi(\beta) = [\sin{\beta} + (\pi - \beta) \cos{\beta}]/\pi$. Then, 
\begin{eqnarray}
\label{eq:pscontrast}
\Csp = \frac{2}{3} \phi(\beta) \left(\frac{R_p}{a}\right)^2 \overline{A(\lambda)},
\end{eqnarray}
where $\overline{A(\lambda)}$ is the wavelength average of the spherical albedo. Assuming Earth-sized planets with $\overline{A(\lambda)}=0.3$, we obtain $\Csp \sim 10^{-8}$ at $a=0.1$ au for the inner edge of the HZ around early M-type stars.

For the $5$-sigma detection of the signal, we obtain
\begin{eqnarray}
\label{eq:snestaf}
\Cinst <  \frac{1}{5^2} \NCCF (\np \Delta \lambda) \Csp.
\end{eqnarray}
Hence, the $\Cinst$ required to detect the signal is estimated as, for instance, $\Cinst \sim 10^{-4} - 10^{-5}$ for $\Csp = 10^{-8} - 10^{-9}$ if $\NCCF$ is hundreds and $\np \Delta \lambda$ is hundreds cts per bin. From those simple estimates, we find that the requirement of the raw contrast are $\Cinst \lesssim 10^{-4}$ for M-type stars. Are these assumptions feasible in the ELT-era ? This is actually an actively developing field and estimating the feasibility is difficult. \citet{2010aoel.confE3007K} simulated the performance of the ExAO for EPICS. Their results show the raw contrast is $\Cinst \sim 10^{-5}$ at 15 mas (the inner edge of HZ for $T_\star$ = 3000 K) and $\Cinst \sim 10^{-6}$ at 100 mas. \citet{2012SPIE.8447E..1XG} considered direct imaging of rocky planets around M dwarfs using ELTs. They presented the PSF raw contrast for a target with an I magnitude of 8.5, assuming wavefront sensing in the I band. They showed that the expected PSF raw contrast in the H-band should be $< 10^{-5}$ for angular separation of $\theta > 5 - 25$ mas, depending on the sampling frequency (Figure 8 in their paper). Hence, we assume that $\Cinst = 10^{-4} - 10^{-5}$ at 15 mas is feasible in the ELT era. 

Because the uncertainty of $\NCCF$ remains unknown in those estimates, depending on the depth of lines, contamination of neighbor lines, and the number of lines that survive against the telluric absorption, we need detailed simulations, as presented in the next section. Strong water lines are generally accompanied by strong absorption of telluric water. As we will see using the simulation, lines of intermediate strength contribute to the total planetary signal.

\subsection{Nighttime airglow and instrumental noise}

In equation (\ref{eq:snestaa}), we neglected the term of the nighttime airglow, $\Ssky$.  The intensity of the nighttime airglow strongly depends on the observing band. We compare the intensity of the nighttime airglow $\Isky = \Ssky \Delta \Omega$ with the speckle intensity $I_\speckle$. If we assume an aperture with a radius $\theta=\lambda/D$ and $D=30$ m, the median, mean and maximum magnitude of the airglow are, respectively, 28,25,19, for the J band, and 27,21,15 for the H band. Because the intensity of the speckle is a factor of $10^3 -- 10^4$ greater than that of the planetary flux for our case, the photon noise of most nighttime airglow lines is negligible. However, the strong airglow lines in the J and H bands may exceed the speckle noise in photon counts at the wavelength of these lines. Fortunately, those strong lines are few in number, so we can mask them when analyzing the spectra. In the next section, we include the nighttime airglow in the simulations to examine its effect.

Instrumental noise, especially readout noise and dark current, depends on the type of detector we assume. For water detection, if we assume that the speckle noise is a factor of $10^3 -- 10^4$ greater than that of the planet flux, we obtain $\sim$ 500 counts per spectral bin ($R \sim 50,000$) and 10 min for a 30-m telescope and 10\% efficiency. Photon noise of the readout and the dark current should be significantly smaller than this value. In this paper, we ignore any detector noise.


\subsection{Stellar Lines as a False Positive}
Late-type stars have intrinsic water lines in their spectra. These lines are imprinted in both the speckles (the speckle stellar lines) and in the planetary scattered light (as the scattered stellar lines). The former can be separated by using the relative velocity of the star and the planet, whereas the latter has the same velocity with respect to the planet. We will examine this effect in Section 4.2.

\section{Cross-Correlation Analysis of the Simulated NIR Spectra of the ELPs}

\subsection{Mock planetary systems and Observational Configuration }

\begin{figure}[htb]
  \includegraphics[width=1.0\linewidth]{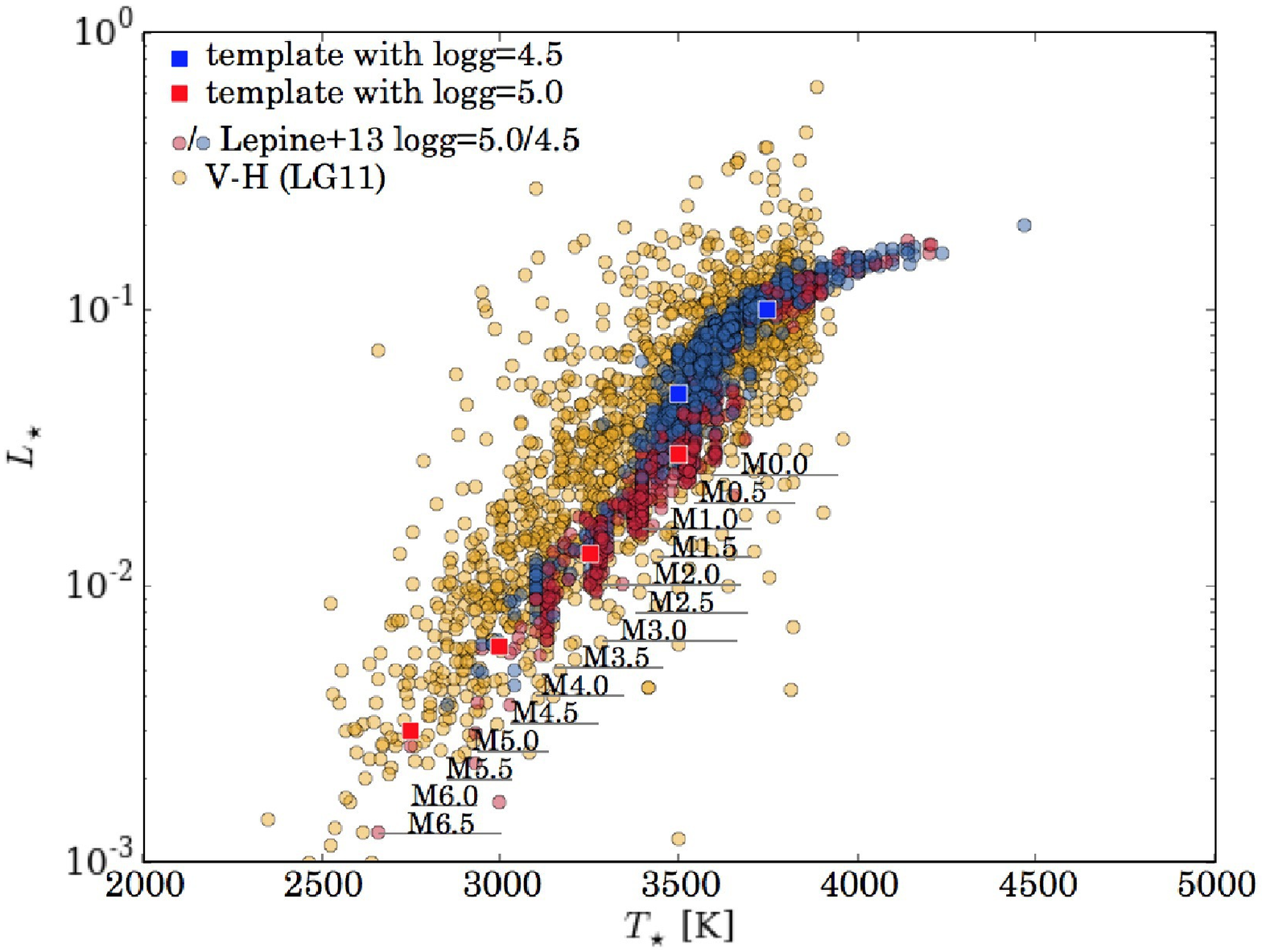}
\caption{ Stellar properties of the nearby late-type stars on the $T_\star$-$L_\star$ plane. The six representative models are shown by squares for logg=4.5 (blue) and logg=5.0 (red). The blue and red circles indicate stars in the spectroscopic catalog of the brightest ($J<9$) M-dwarfs in the northern sky by \citet{2013AJ....145..102L}. The stellar luminosity $L_\star$ is estimated from J,H,and K magnitude.  The ranges of spectral type for \citet{2013AJ....145..102L} are shown by horizontal bars. The yellow circles are from the photometric catalog of the all-sky late-type stars with $J<10$ by \citet{2011AJ....142..138L}. The stellar temperature is estimated by adopting the empirical temperature-(V-H) relation \citep{2008MNRAS.389..585C}.
\label{fig:mystar}}
\end{figure}

We determine the stellar parameters (stellar luminosity, effective temperature and logg) based on the spectroscopic catalogs of nearby late-type stars. Figure \ref{fig:mystar} shows the $T_\star$-$L_\star$ relation of nearby late-type stars based on \citet{2013AJ....145..102L,2011AJ....142..138L}. We pick up six representative points on the plane shown by squares. The parameters of these six models are summarized in Table \ref{tab:msys}. Among the six models, we use the stellar model with $T_\star = 3000$ K as a fiducial model. For the $T_\star=3000$ K model, the corresponding HZ is 0.08 - 0.16 au \citep{2013ApJ...765..131K}. We decide to use the inner edge as is the case for the Earth, a=0.08 au, as the semimajor axis of the planet. Then, the planet-star contrast is $\sim 2 \times 10^{-8}$. The maximum angular separation for a circular orbit is 16 mas at 5 pc, corresponding to 1.4--2.6 $\lambda/D$ for 0.95--1.8 $\mu$m with $D=30$ m. We expect that the next-generation high-contrast instruments will significantly reduce the stellar speckles in this angular separation range \citep[e.g.][]{2012SPIE.8447E..1XG}. The planetary system we assume is similar to Kepler-186 system\citep{2014Sci...344..277Q}, GJ 667C \citep{2012ApJ...751L..16A}, and GJ 832c \citep{2014ApJ...791..114W} for a planet within or close to the HZ around M-type stars.

Because the aim of the paper is to see how the method works for the Earth analog, the planet has the Earth radius and the same structure as the terrestrial atmosphere, although the atmospheric structure depends on the stellar type owing to the UV environment \citep{2010AsBio..10..751S} and the tidal locking \citep[e.g.][]{1997Icar..129..450J,2010JAMES...2...13M, 2013ApJ...771L..45Y}. 

The relative velocity of the planetary system to the observer is set to 20 km/s. In reality, this velocity includes the orbital radial velocity of the Earth ($\sim 30$ km/s as the orbital velocity) and the relative velocity between the Sun and the observed system. The relative radial velocity of the planet to the system is set to 30 km/s so as to reproduce the typical radial velocity of the planet in the HZ with the intermediate value of the orbital inclination $i \sim 45^\circ$.

\begin{table}[!htb]
\begin{center}
\caption{Stellar and Planetary Properties of the Mock Systems \sc \label{tab:msys}}
  \begin{tabular}{ccccccc}
   \hline\hline 
   \multicolumn{7}{c}{Host Star} \\
   \hline
   $T_\star$ [K] & 2750 & 3000 & 3250 & 3500 & 3500 & 3750 \\
   $d$ [pc] & \multicolumn{6}{c}{ 5 pc} \\
   logg & 5 & 5 & 5 & 5 & 4.5 & 4.5 \\
   $L_\star/L_\odot$ & 0.003 & 0.006 & 0.013& 0.03& 0.05& 0.1 \\ 
   $R_\star/R_\odot$ & 0.239 & 0.285 & 0.363 & 0.476 & 0.614 & 0.756 \\
   metalicity  &    \multicolumn{6}{c}{solar metalicity} \\
   $\Vsys$ & \multicolumn{6}{c}{20 km/s} \\
   \hline
   \multicolumn{7}{c}{Planet in the HZ} \\
   \hline
   $\Vp$ & \multicolumn{6}{c}{30 km/s}\\
   $R_p$ & \multicolumn{6}{c}{$R_\oplus$} \\   
   $a$ & 0.06 & 0.08 & 0.12 & 0.19 & 0.24 & 0.34 \\
   $\theta$ [mas] $\,^\dagger$ & 12 & 16 & 24 & 38 & 48 & 68 \\
\multicolumn{3}{l}{$\dagger$: the maximum angular separation. } \\
\end{tabular}
\end{center}
\end{table}

The instrumental configuration is summarized in Table \ref{tab:obsset}. We assume a next-generation telescope with a diameter of 30 m. The total throughput of the high-dispersion spectrograph is generally lower than the photometric detector. We assume 10 \% of the total throughput with a spectral resolving power $R=50,000$.

\begin{table}[!htb]
\begin{center}
\caption{Instrumental Configuration \label{tab:obsset}}
  \begin{tabular}{ccc}
   \hline\hline
    & symbol & value\\
   \hline
 Telescope diameter & $D$ & 30 m \\
 Total throughput & $\Thput$ & 0.1 \\
 Resolving power & $R$ & 50,000  \\
 \multicolumn{2}{c}{Band for water detection} & 0.95-1.8 $\mu$m (Y,J,H)\\
 \multicolumn{2}{c}{Band for oxygen detection} & 1.27 $\mu$m \\
 \multicolumn{2}{c}{Band for the scattered stellar lines} & 0.95-1.8 $\mu$m (Y,J,H)\\
 Raw contrast & $\Cinst$ & $10^{-4}$,$10^{-5}$ \\
 Exposure time & $\Texp$  & 10 days \\
\end{tabular}
\end{center}
\end{table}

\subsection{Radiative Transfer}

To simulate both the scattered spectra of the ELP and the atmospheric transmittance of our Earth, we use the radiative transfer code libRadtran\footnote{http://www.libradtran.org} \citep{mayer2005technical} with the line-by-line scheme (LBL). The LBL optical depth is computed by using Py4CATS\footnote{See the website of libradtran for the detail.} from the line parameter database HITRAN2012 \citep{rothman1999hitran}. The LBL optical depth includes the molecular lines of water ($\mathrm{H_2O}$), oxygen  ($\mathrm{O_2}$), carbon dioxide  ($\mathrm{CO_2}$) and nitrous oxide ($\mathrm{N_2O}$).  The AFGL atmospheric constituent profile ( U.S. standard atmosphere 1976) is used for the atmospheric structure.

We use the stellar spectral models generated by Phoenix \citep{2013A&A...553A...6H} as the input stellar input flux. Interpolating the templates on the temperature grid, we create the unbroadened stellar spectrum for a given set of parameters. Data sampling of the stellar spectra is $\Delta \lambda = 0.002$ nm  ($R \sim 5 \times {10}^5$ at 1 $\mu$m), which is sufficiently high for our purpose. The synthetic spectra are smoothed according to stellar rotation and macroturbulence broadening kernel (see. e.g. \cite{2011ApJ...742...69H}). We assume the macroturbulent velocity of 1.5 km/s. Based on the observational results of the projected rotational velocity of nearby M stars \cite[e.g.][]{2012AJ....143...93R,2003ApJ...583..451M}, we adopt $\Vrot=5$ km/s as a fiducial value until \S \ref{sec:ssl}.

The planetary flux is directly derived from radiance computed by using radiative transfer. To reduce the computation time, we use a representative geometry, instead of considering all of the facets on the planetary surface, that has a solar-zenith angle of $60^\circ$, , azimuth angles of ingoing and outgoing rays of $180^\circ$,  and a view zenith angle of $30^\circ$, roughly corresponding to the central position of the visible and illuminated area of the planet when the star-planet-observer angle is $\beta = 90^\circ$. We computed both the clear-sky spectrum and the cloudy-sky spectrum and took an average of them (with the cloud fraction being 50\%). We assume water clouds with an optical depth of 20 and an altitude of 2-4 km. The ground albedo was set to 0.1 so as to reproduce the planetary albedo of our Earth, $A \sim 0.3$.

The transmittance of our Earth at a 4-km altitude from sea level is also computed by using the same setting of radiative transfer as for the clear-sky case. Given the long exposure of the target, we use $45^\circ$ as the representative altitude of the observation. The simulated nighttime airglow spectrum is taken from the Gemini website (Lord, S.D. 1992, NASA Tech. Memo. 103957 and Gemini Observatory) \footnote{http://www.gemini.edu/sciops/telescopes-and-sites/observing-condition-constraints/ir-background-spectra}. We use the Mauna Kea sky emission with airmass = 1.5 and water vapor column = 1.0 mm.

\begin{figure}[htb]
  \includegraphics[width=1.0\linewidth]{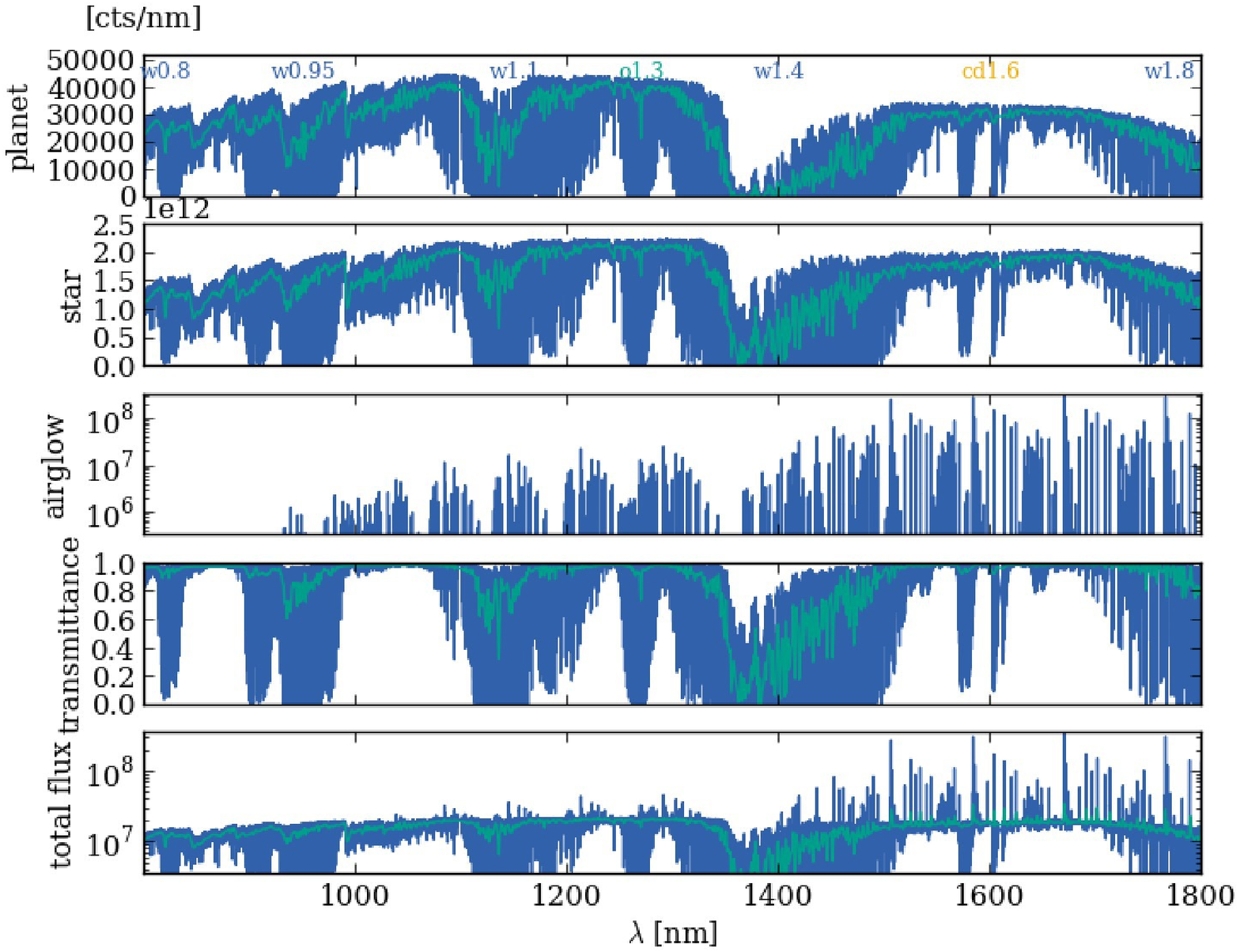}
\caption{ Examples of simulated NIR spectra. From top to bottom, $F_\mathrm{p} (\lambda)$ and $F_\star (\lambda)$, the nighttime airglow, the transmittance $T(\lambda)$ at the 4 km/s altitude where the telescope is located, and the total spectrum, $f_\tot$ for $\Cinst=10^{-5}$. The spectral resolution is $\Delta \lambda=0.002$ nm (blue). The green lines are smoothed by binning with a width of 1 nm. The labels w, o, and cd on the top panel indicate strong features from $\mathrm{H_2 O}$,$\mathrm{O_2}$, and $\mathrm{C O_2}$, respectively. \label{fig:mixspectra}}
\end{figure}

Figure \ref{fig:mixspectra} shows an example of the simulated spectra. We combine the planetary spectra, the stellar spectra, the transmittance of our Earth, and the airglow according to equations (\ref{eq:totalflux}), (\ref{eq:planet}), and (\ref{eq:speckle}) as shown in the bottom panel of Figure \ref{fig:mixspectra}. The instrumental noise $f_n$ is neglected. The spectra are binned with a spectral resolution of $R = 50,000$. The expected number of photons for each bin is computed based on the telescope diameter of $D$, the exposure time $\Texp$, and the total throughput $\Thput$. We added shot noise to the spectra according to a Poisson distribution.

\begin{figure}[htb]
  \includegraphics[width=1.0\linewidth]{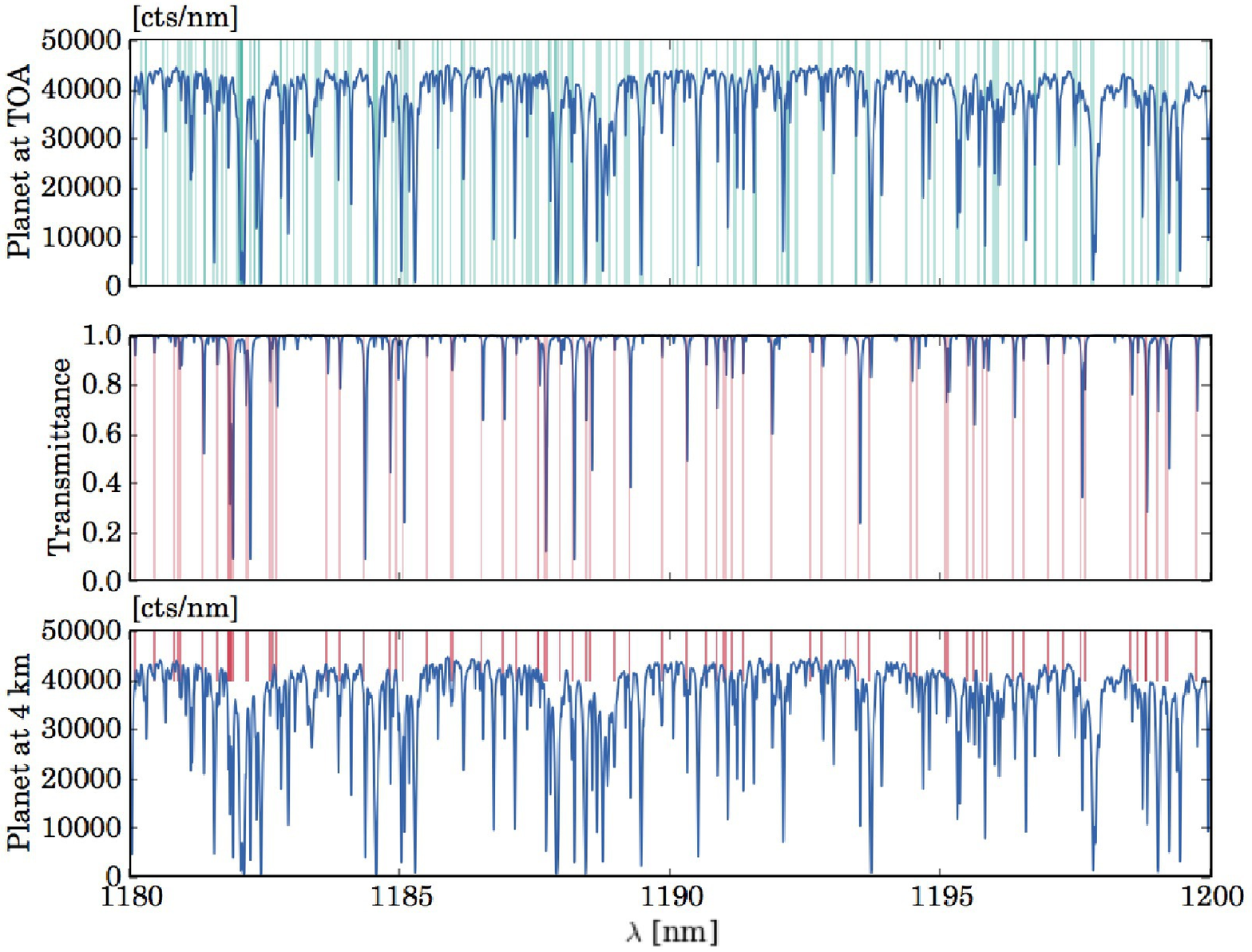}
\caption{Difference between the Doppler-shifted water lines and the transmittance of our Earth. From top to bottom are the planetary spectrum $f_p (\lambda)$, the transmittance of our Earth at the 4-km altitude where the telescope is located, $T(\lambda)$, and the planetary spectrum at the 4 km altitude, $T(\lambda) \, f_p (\lambda)$. The green and red vertical lines indicate the positions of water vapor lines with line strength $> 10^{-24} \mathrm{cm^{-1}/cm^{-2}}$ in the Doppler-shifted frame ($v=\Vp +\Vsys=50$ km/s) and in the rest frame ($v=0$ km/s). 
\label{fig:mixspectra_water}}
\end{figure}

Before analyzing the mock spectra, we demonstrate how the planetary lines are separated from the telluric absorption lines by using this simulation. Figure \ref{fig:mixspectra_water} shows the unabsorbed planetary spectrum, the transmittance of our Earth, and the planetary spectrum after transmission through the atmosphere of our Earth. As shown by the green and red vertical lines that indicate the wavelength of the strong water lines in the Doppler-shifted frame and in the rest frame, respectively, the water vapor lines from the exoplanets after the transmittance of our Earth can be separated from the telluric one if the transmittance of our Earth is not very large. High-dispersion spectroscopy has the same virtue for other terrestrial biosignatures, such as oxygen and carbon dioxide.

\subsection{Data Reduction and Cross-Correlation Function Analysis}
Both the transmittance spectrum of our Earth and the speckle spectrum are deducted from the combined spectra to increase the signal from the planetary light. In principle, those spectra can be estimated by the simultaneous observation of the speckles at no planet position and the sophisticated method to fit the transmittance of our Earth. There can be several possible sources of systematics in that procedure, i.e., uncertainty of the stellar speckle spectrum, the nighttime airglow, and the transmittance of our Earth. In this paper, we mainly focus on the most optimistic case, the photon noise limit: We know the accurate speckle stellar spectra, and the transmittance of our Earth can be perfectly corrected. For the photon noise limit, the signal-to-noise ratio S/N of the planetary spectrum is purely determined by the photon counts and there is no systematic noise in the planetary spectrum. In Section 3.7, however, we consider the sensitivity of the results to the systematics.

\begin{figure}[htb]
  \includegraphics[width=1.0\linewidth]{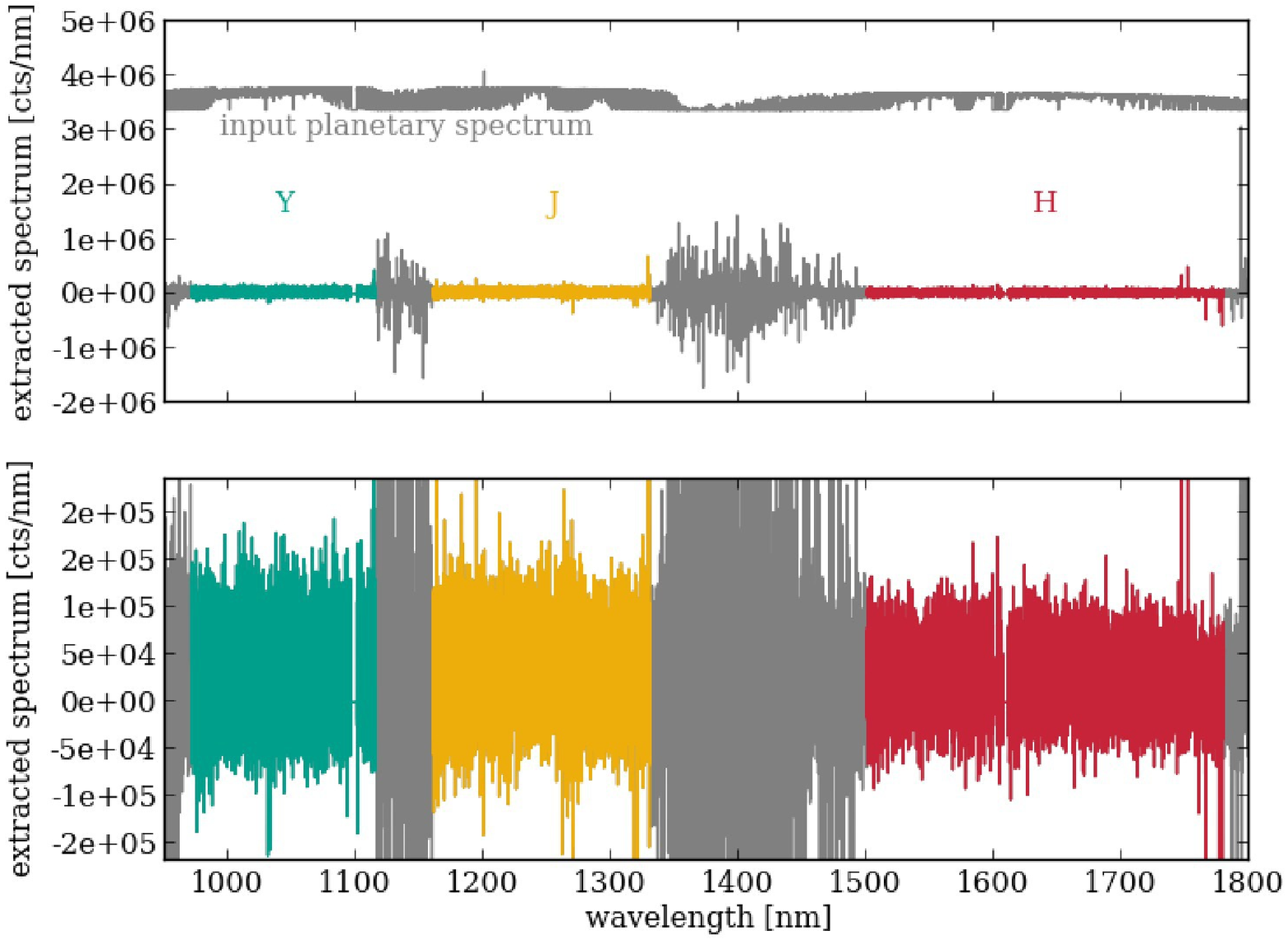}
\caption{Extracted planetary spectrum from the integrated spectrum ($\Cinst=10^{-5}$). The colored regions are used for the cross-correlation analysis and the gray regions are masked to eliminate the regions with large photon noises. For reference, the input planetary spectrum is shown by the gray line over the signal. The bottom panel displays an enlarged view of the top one. \label{fig:spectra}}
\end{figure}

To distinguish the mock spectra with photon noise from the theoretical spectra with no noise, we use $\Fobs$ for the mock spectra instead of $f$. The extracted planetary spectrum for the photon noise limit $\Fpesti(\lambda)$ is expressed as
\begin{eqnarray}
\label{eq:ideal}
\Fpesti(\lambda) = \frac{\Fobs_\mathrm{tot}(\lambda)  - f_\speckle(\lambda) - \fsky (\lambda) }{T(\lambda)}.
\end{eqnarray}
Figure \ref{fig:spectra} shows an example of $\Fpesti(\lambda)$ with the input planetary spectrum $f_p(\lambda)$ (gray). It is impossible to discern any features of the planetary spectrum by eye. We extract the signal by using a cross-correlation analysis, a successful means to derive the molecular features in hot Jupiters as reviewed in the Introduction. The normalized cross-correlation function (CCF) is defined by
\begin{eqnarray}
\label{eq:ccf}
\ccp \star \Fpest (v) &\equiv& \int d v^\prime \frac{[\ccp(v^\prime-v) - \overline{\ccp}] [\Fpest(v^\prime) - \overline{\Fpest}]}{\sigma(\ccp) \sigma(\Fpest)}, 
\end{eqnarray}
where $\ccp (v)$ is the spectral template of molecules or the stellar lines whose features we want to extract from the spectrum, $\overline{\ccp}$ and $\overline{\Fpest}$ are the average of $\ccp$ and $\Fpest$, and $\sigma(\ccp)$ and $\sigma(\Fpest)$ are the standard deviations of $\ccp$ and $\Fpest$. Before computing the CCF, we exclude the 5 $\sigma$ outliers from $\Fpest$. This clipping procedure is important when $\Fpest$ has bins with large error resulting from the strong telluric absorption lines.

\section{Detectability of Molecules and Its False Positive}
\subsection{Water Vapor}

In the NIR band, there are numerous water vapor lines over a wide range of line strength. The most effective way to extract the molecular signal is to use the template spectrum adequately simulated by radiative transfer. However, in general, we do not utilize any information on the atmospheric structure of the exoplanet. We try two types of spectral template: a binary model and an exponential model. 

\begin{figure}[htb]
  \includegraphics[width=1.0\linewidth]{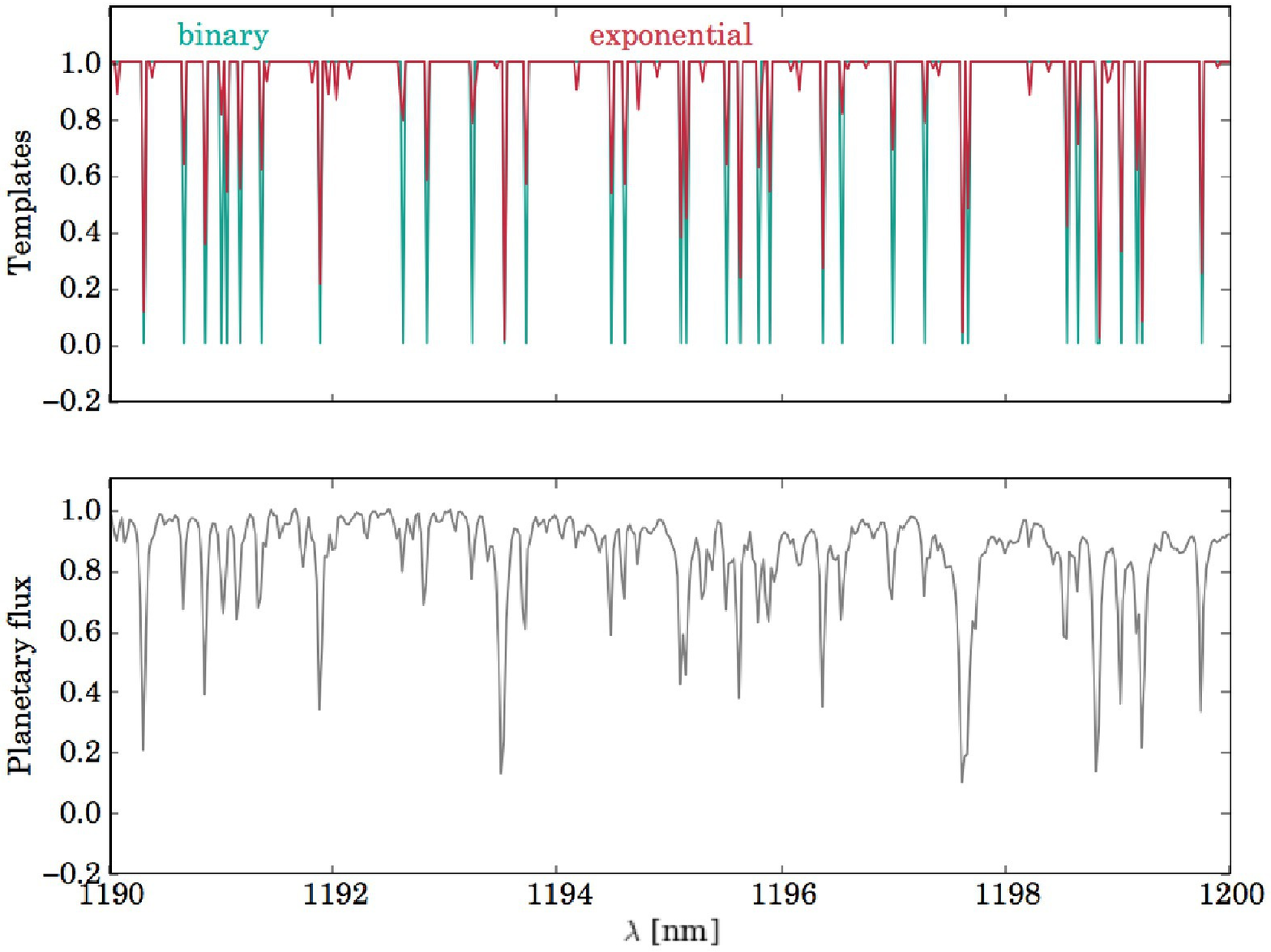}
\caption{Spectral templates of water for the CCF analysis. The top panel shows the binary model with $\ulim=10^{-24} \, \mathrm{cm^{-1}/cm^{-2}}$ and the exponential model with $u_c = 5 \times 10^{-24} \, \mathrm{cm^{-1}/cm^{-2}}$. The bottom panel is the normalized planetary spectrum of the ELP. \label{fig:ccp}}
\end{figure}

In the binary model, for the $i$-th molecular line with line strength $u_i(\lambda_i) \ge \ulim$, $\ccp(\lambda_i)=0$, otherwise $\ccp(\lambda_i)=1$. The spectral template of the binary model is created based on HITRAN2012. Water vapor has an enormous number of molecular lines in the visible and infrared bands. If the binary template includes all lines, the CCF signal becomes very small. Hence we must choose an adequate criterion for the line strength. We find that $\ulim = 10^{-24} \, \mathrm{cm^{-1}/cm^{-2}}$ covers most strong lines of scattered light (Figure \ref{fig:ccp}). Because the typical width of water line is close to the bin size of the spectral resolution of $R=50,000$, we do not smooth the template.

The exponential template is based on the relation between transmissivity and line strength:
\begin{eqnarray}
g_t (\lambda) =   \left\{
\begin{array}{lc}
e^{- u_i/u_c} & \mbox{for $\lambda = \lambda_i$}\\
1 & \mbox{elsewhere, } \\
\end{array}
\right.
\end{eqnarray}
where the critical line strength, $u_c$ represents the saturation level of the lines. As shown in Figure \ref{fig:ccp}, the exoponetial model with  $u_c= 5 \times 10^{-24} \, \mathrm{cm^{-1}/cm^{-2}}$ gives a better approximation of the scattetered spectra of the ELP. Those adequate criterions for the line strength are related to the column density of water. The AFGL atmospheric constituent profile has the water column density of $N_\mathrm{H_2 O} \approx 5 \times 10^{22} \mathrm{cm^{-2}}$. The air-broadend half width is $\Delta \nu \sim 0.1 \mathrm{cm^{-1}}$. The criterion of the line strength, $u$, should satisfy $\tau \sim u N_\mathrm{H_2 O}/\Delta v \sim 1$. Hence, the adequate criterions of $\ulim$ and $u_c$ are close to $\Delta \nu/N_\mathrm{H_2 O} \sim 2 \times 10^{-24} \mathrm{cm^{-1}/cm^{-2}}$. In reality, the column density of water of the ELP is unknown, and $\ulim$ or $u_c$ is regarded as a free parameter to search for the CCF signal.

To mask the wavelength regions that exhibit high photon noises, we exclude the ranges where the telluric extinction is high (the gray regions in Figure \ref{fig:spectra}). As the result of this masking procedure, most water lines with line strength $> 10^{-22} \, \mathrm{cm^{-1}/cm^{-2}}$ are eliminated. We confirmed that this procedure is crucial to increase the signal-to-noise ratio. Hence, this method mainly uses water vapor lines of intermediate strength  ($10^{-24} - 10^{-22} \, \mathrm{cm^{-1}/cm^{-2}}$).

\begin{figure}
\begin{center}
  \includegraphics[width=\linewidth]{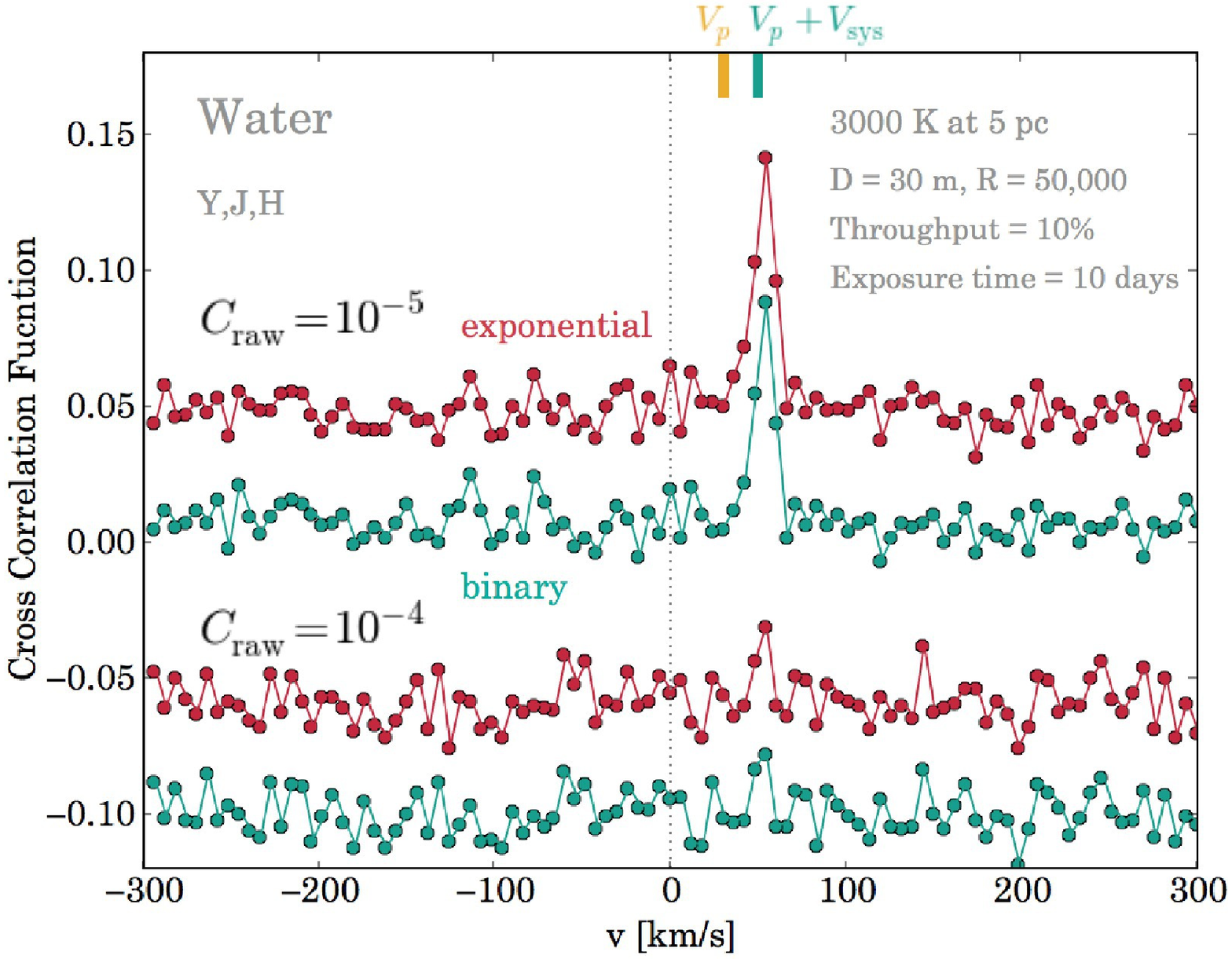}
\caption{CCFs of the water vapor to the simulated high-resolution spectra of the ELP around M-type stars ($T_\star=3000$ K). We consider two cases of the raw contrast: $\Cinst=10^{-4}$ (bottom) and $\Cinst=10^{-5}$ (top).  The spectra within 0.95-1.8 $\mu$m (Y+J+H) are analyzed. The color indicates the difference of the spectral template, corresponding to the binary model with $\ulim = 10^{-24} \mathrm{cm^{-1}/cm^{-2}}$ (green) and the exponential model with $u_c = 5 \times 10^{-24} \, \mathrm{cm^{-1}/cm^{-2}}$ (red)  . The upper red and lower red and green curves are artificially shifted for clarity of presentation.
\label{fig:vwater}}
\end{center}
\end{figure}

Figure \ref{fig:vwater} displays the CCFs between the spectral templates of water and the mock spectra for $T_\star=3000$ K. The CCFs exhibit the clear feature of Doppler-shifted water lines at $v=\Vp+\Vsys$ at least for $\Cinst = 10^{-5}$, whereas the signal for $\Cinst = 10^{-4}$ is marginal. The binary model and the exponential model provide almost the same results. We use the exopnetial model in the rest of the paper.

Because the peak has $\sim$ three bins of the width, we take the average of the CCF over each bin and compute the standard deviation between -300 to 300 km/s excluding the peak. Then we define the signal-to-noise ratio by the ratio of the peak (the mean of the three bins around $v=50$ km/s) and the standard deviation of these averaged data. We find 3 $\sigma$ and 16 $\sigma$ detections for $\Cinst = 10^{-4}$ and $10^{-5}$, respectively, for the exopnential model. As expected from equation (\ref{eq:snestaa}), S/N $ \propto 1/\sqrt{\Cinst}$. Though there are $\sim$ 848 lines after the 5 $\sigma$ clipping with line strength $ > 10^{-24} \, \mathrm{cm^{-1}/cm^{-2}}$ in the Y-H bands, we obtain $\NCCF \sim 270$ using equation (\ref{eq:snestaa}). There are several possible explanations of the decrease of the effective number: (1) The water lines with strength $> 10^{-24} \, \mathrm{cm^{-1}/cm^{-2}}$ are from various depths and some of the lines have a shallower line contrast than that of $\Csp$ (see Figure \ref{fig:mixspectra_water}). (2) Neighboring lines contaminate the signal and increase the noise. (3) There are coincidental telluric absorption lines (in spite of the exclusion of the strongly absorbed range from the analysis).

\begin{figure}
\begin{center}
  \includegraphics[width=\linewidth]{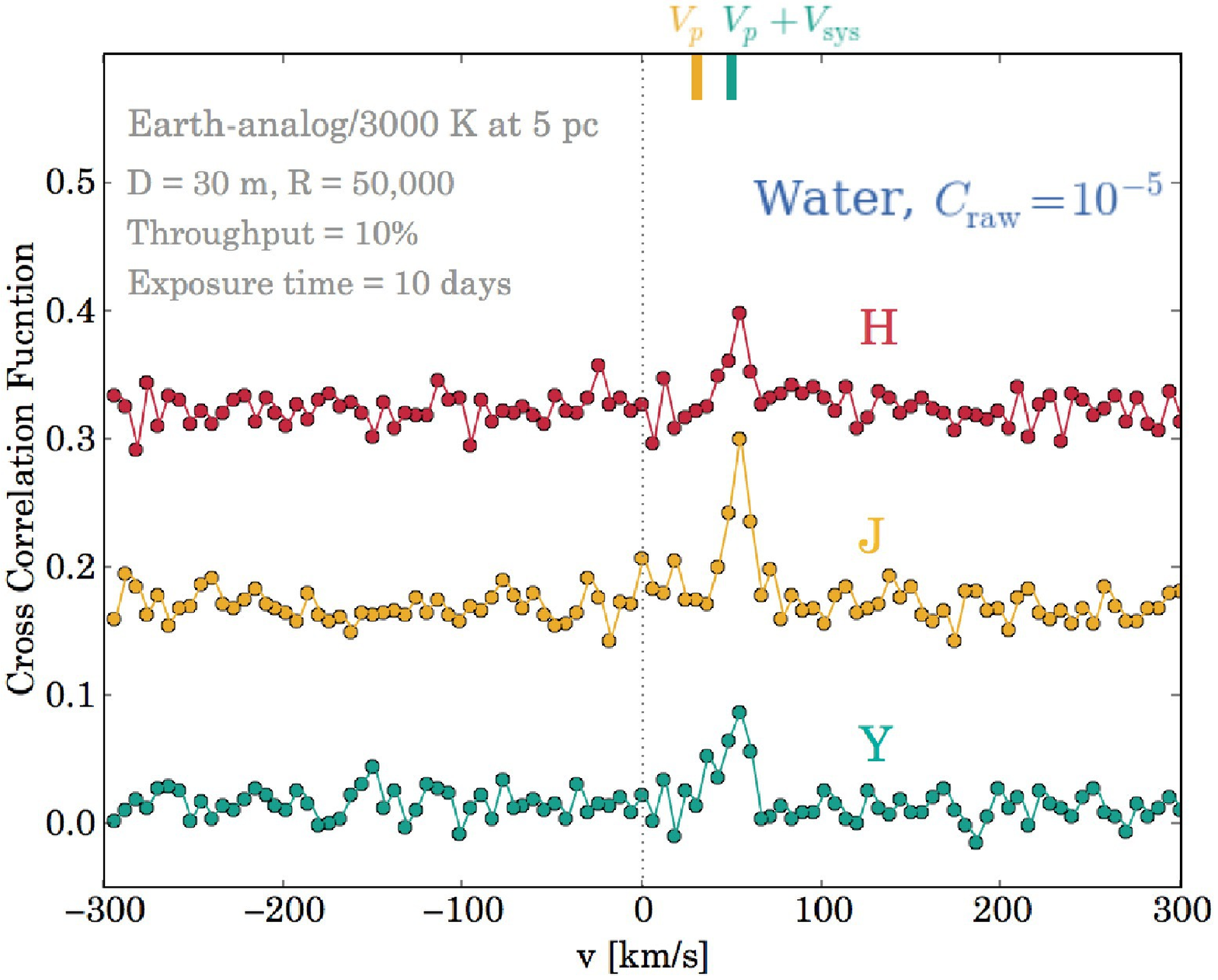}
\caption{Dependence on the bands used for the CCF analysis of water vapor (for stars with $T_\star = 3000$ K). From bottom to top, Y,J,and H bands are used. In this figure a raw contrast of $\Cinst=10^{-5}$ is assumed. Curves corresponding to Y and H bands are artificially shifted in the $y-$ direction for clarity of presentation. \label{fig:color}}
\end{center}
\end{figure}

We also examine the strength of the signal for each band. In the NIR band we consider, there are three main wavelength ranges corresponding to Y, J, and H bands, as indicated by colors in Figure \ref{fig:spectra}. Figure \ref{fig:color} shows the color dependence of the signal with the same conditions as in Figure \ref{fig:vwater} (the exopnential model and $\Cinst=10^{-5}$). Because strong water lines exist over all the NIR bands, all of the Y,J and H bands contribute to the water vapor signal. Among those bands, the J band exhibits the strongest feature. This fact is notable because the J band also has a strong oxygen feature.

\begin{figure}
\begin{center}
  \includegraphics[width=\linewidth]{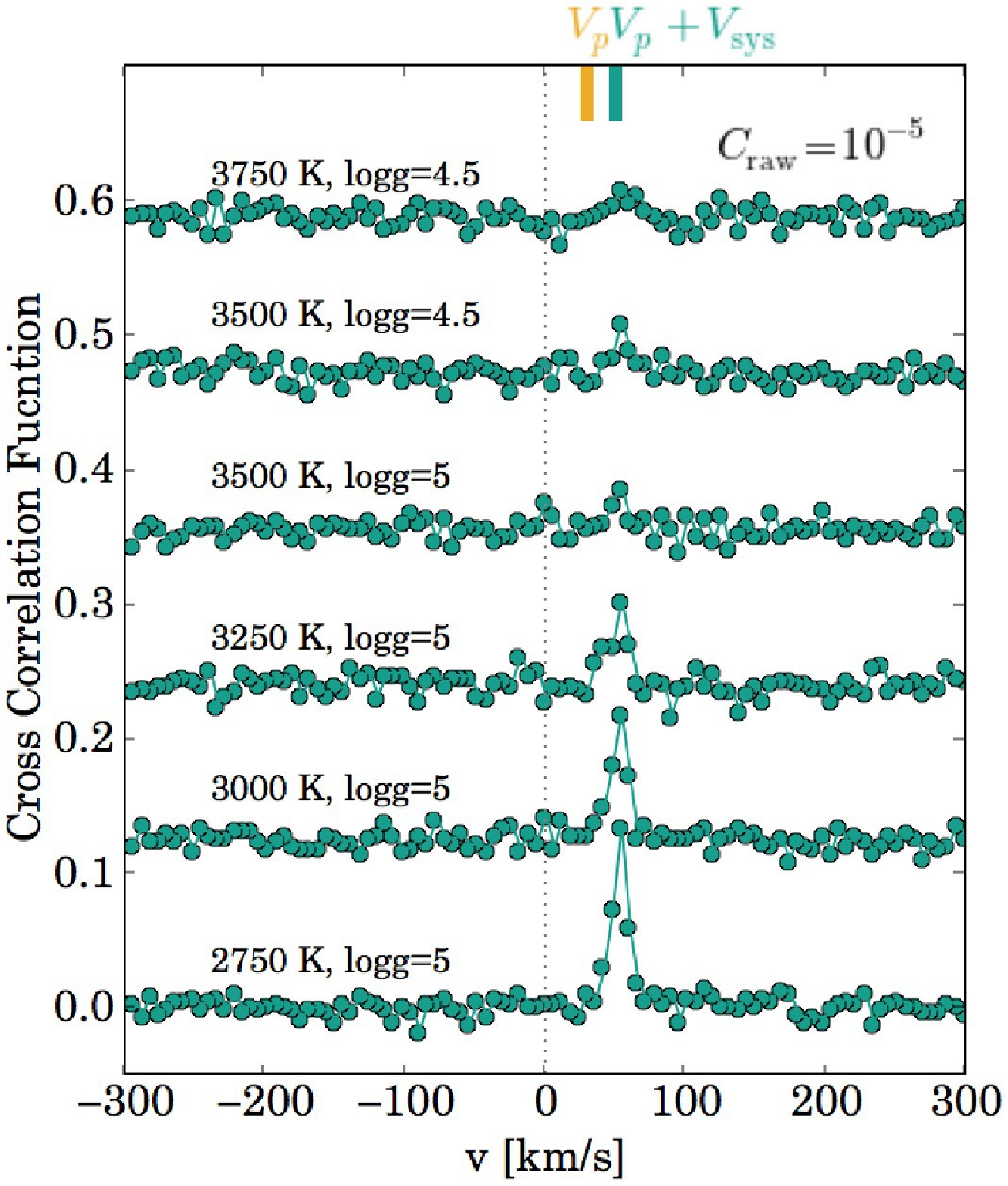}
\caption{Dependence of the water signal on the stellar models ($T_\star = 2750$ K to 3750 K). We assume $\Cinst=10^{-5}$. \label{fig:ccpTeff}}
\end{center}
\end{figure}

Figure \ref{fig:ccpTeff} shows the CCFs for the six representative stellar models. The water signal increases with decreasing $T_\star$ for a constant instrumental contrast of $\Cinst = 10^{-5}$ because the planet-star contrast at the inner edge of the HZ also increases. In reality, $\Cinst$ becomes worse as the angular separation decreases. We did not include this effect because it depends on the details of the instrument. Instead, we conclude that instruments having $\Cinst = 10^{-5}$ can detect water signals from ELP for $T_\star \lesssim 3500 $ K.

\subsection{Intrinsic water lines in the stellar spectra \label{sec:iwl}}

The intrinsic water lines in the speckle stellar lines can be separated by using the relative velocity of the star and the planet, whereas the water lines in the scattered stellar lines (see Table 1) has the same velocity with respect to the planet. The latter can cause false positives. 

\begin{figure}
\begin{center}
  \includegraphics[width=\linewidth]{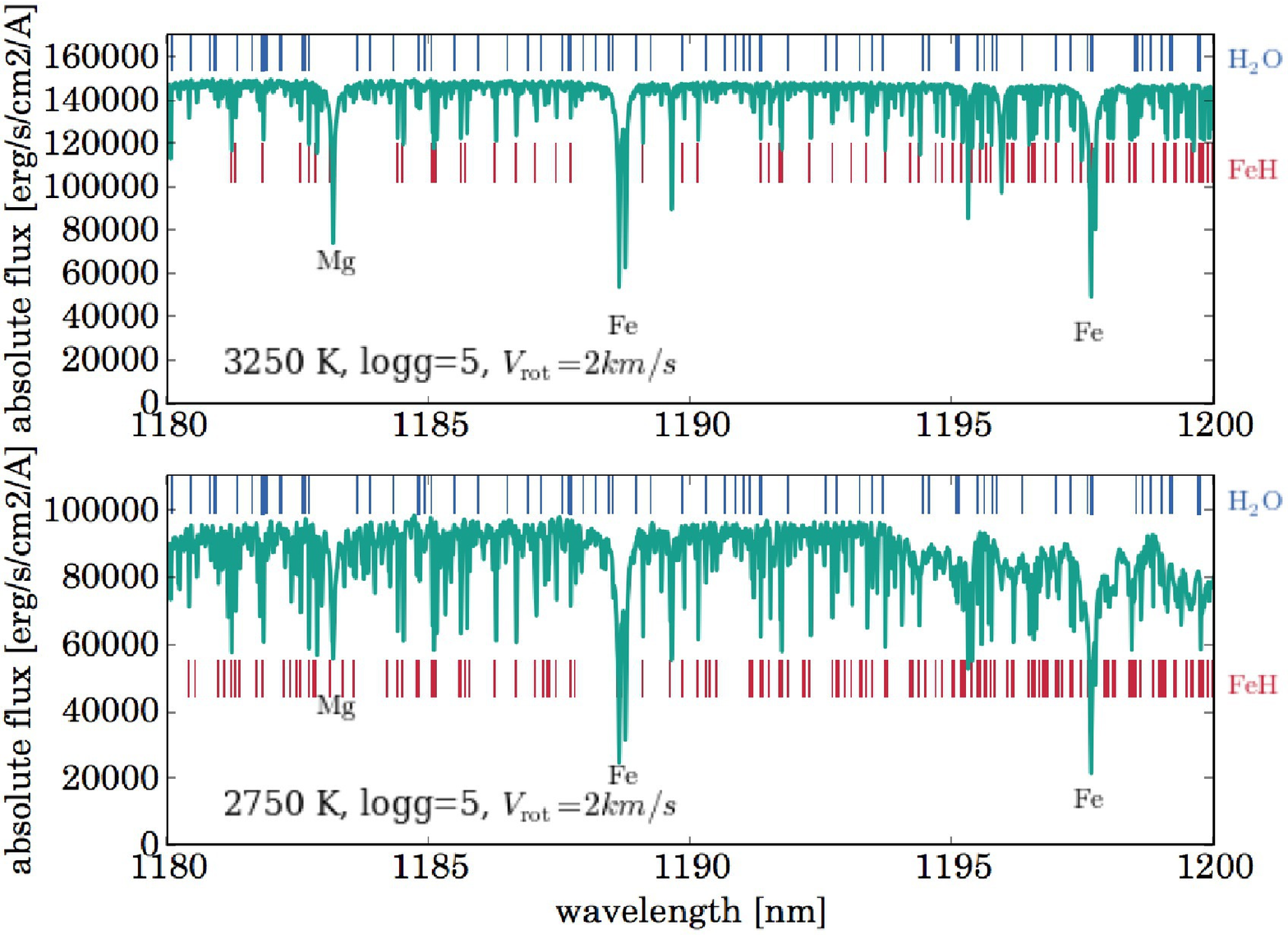}
\caption{Phoenix stellar spectra (1080-1200 nm) for $T_\star=3250$ and 2750 K. FeH lines are identified by red bars, which are identified by referring to \cite{2010A&A...523A..58W}. Blue bars are water lines with line strength $>10^{-24} \mathrm{cm^{-1}/cm^{-2}}$. \label{fig:spentraJ}}
\end{center}
\end{figure}

Figure \ref{fig:spentraJ} shows parts of the Phoenix stellar spectra for $T_\star$=3250 K and 2750 K. There are numerous FeH lines (red bars) and several metal lines in this wavelength range, whereas there are no strong water lines, as shown by disagreement between the blue bars (water vapor with line strength $> 10^{-24} \mathrm{cm^{-1}/cm^{-2}}$) and the features of the spectra. This demonstrates that the strong water vapor signal from ELPs is not affected by the intrinsic water vapor lines of the stellar spectra. We note that the Phoenix stellar spectrum for $T_\star=2750 $K exhibits intrinsic water line features, for instance, in 1340-1440 nm. However, we do not use this range for analysis because strong telluric absorption lines also exist in this range. 

\begin{figure}
\begin{center}
  \includegraphics[width=\linewidth]{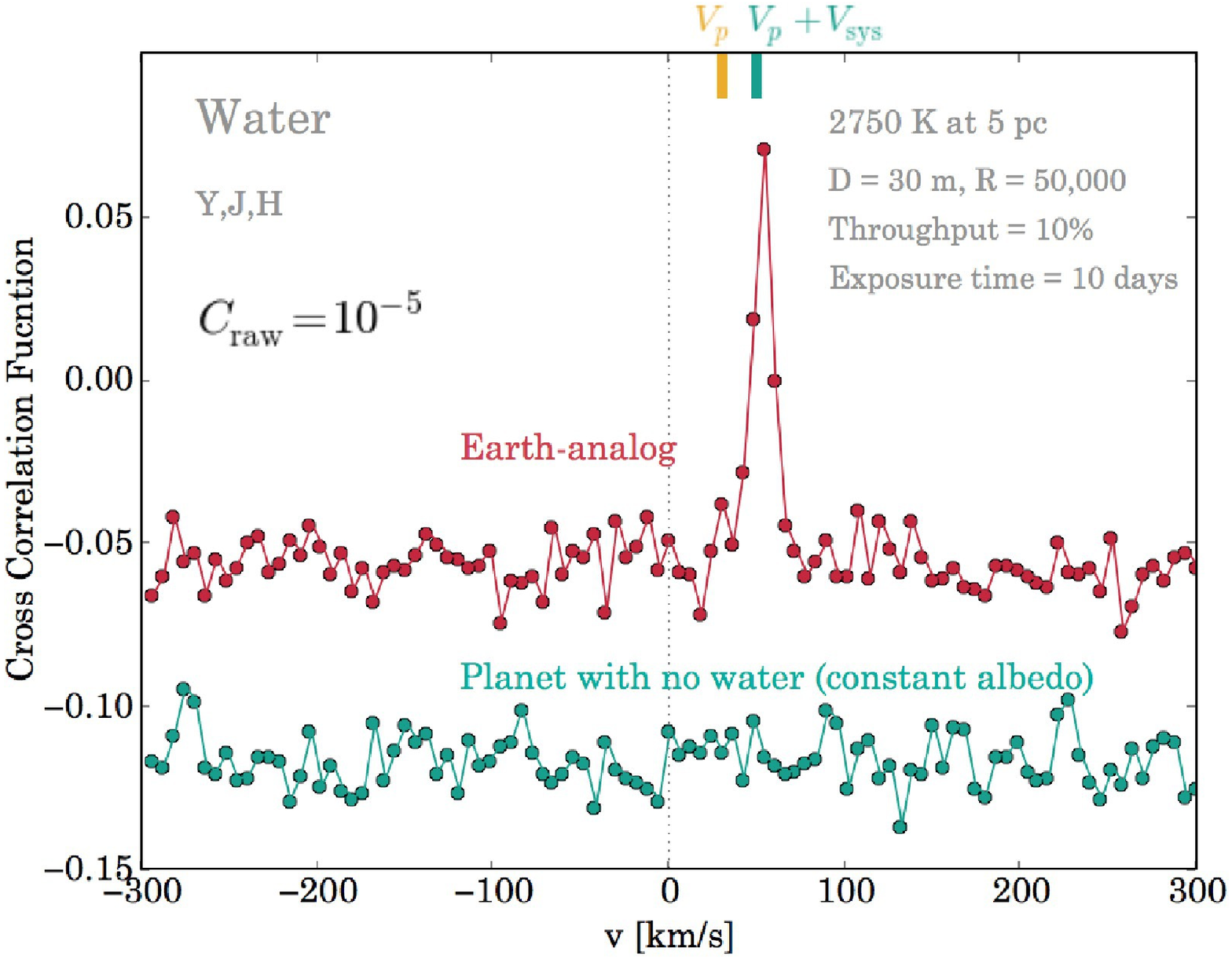}
\caption{CCF analysis of water for a planet having constant reflectivity (no water; green curve). For reference, we show the same analysis for the Earth analog (with water; red curve). We adopt a stellar spectra with $T_\star = 2750$ K and $\Vrot=2$ km/s. \label{fig:flessalbedo}}
\end{center}
\end{figure}

To confirm that the stellar intrinsic water lines do not affect our analysis, we also perform a CCF analysis for a planet having featureless albedo (a constant $A(\lambda)$). If the stellar intrinsic water lines are sufficiently strong to contaminate the planetary lines as the scattered stellar lines, the CCF signal of water should be detected even for the featureless albedo (see equation 9 and Table 1) . As expected, we do not detect the CCF signal of water for this case (Figure \ref{fig:flessalbedo}). Hence, we conclude that the stellar water lines do not produce a false positive water detection signal for the Phoenix synthetic spectra. However, checking whether or not the stellar spectrum has water lines in individual cases is important.

\subsection{Sensitivity to Systematics \label{sec:sys}}

So far, we have assumed that we can perfectly estimate the stellar speckle, the transmission of our Earth, and the nighttime airglow. Here, we investigate the sensitivity of the result to the accuracy of those estimates. Changing the true spectrum $f_\speckle(\lambda)$ and the true transmission $T(\lambda)$ to inaccurate ones $f^{\prime}_\speckle(\lambda)$  and $T^\prime(\lambda)$ in equation (\ref{eq:ideal}),
\begin{eqnarray}
\Fpest(\lambda) = \frac{\Fobs_\mathrm{tot}(\lambda)  - f^{\prime}_\speckle(\lambda) - \fsky (\lambda) }{T^\prime(\lambda)},
\end{eqnarray}
we examine the sensitivity of the results to the speckle subtraction and the telluric correction.

 The degree of the accuracy for the speckle subtraction is characterized by the deviation of the used spectrum from the true spectrum, 
\begin{eqnarray}
\Delta f_\speckle \equiv f_\speckle(\lambda) - f^{\prime}_\speckle(\lambda).
\end{eqnarray}
In general, stronger telluric lines are more difficult to fit. To include this effect, we compute $T^\prime (\lambda_i)$ at the telluric water absoption lines $\lambda = \lambda_i$ as
\begin{eqnarray}
r_T \equiv \frac{1 - T^\prime (\lambda_i)}{1 - T(\lambda_i)}
\end{eqnarray}
where $r_T$ is the ratio of the estimated depth of the telluric lines to the true depth. We adopt $r_T=0.95$, which means that the lines with depth $d$ ($0<d<1$) are underestimated by $5 \times d$ \%. 

\begin{figure}
\begin{center}
  \includegraphics[width=\linewidth]{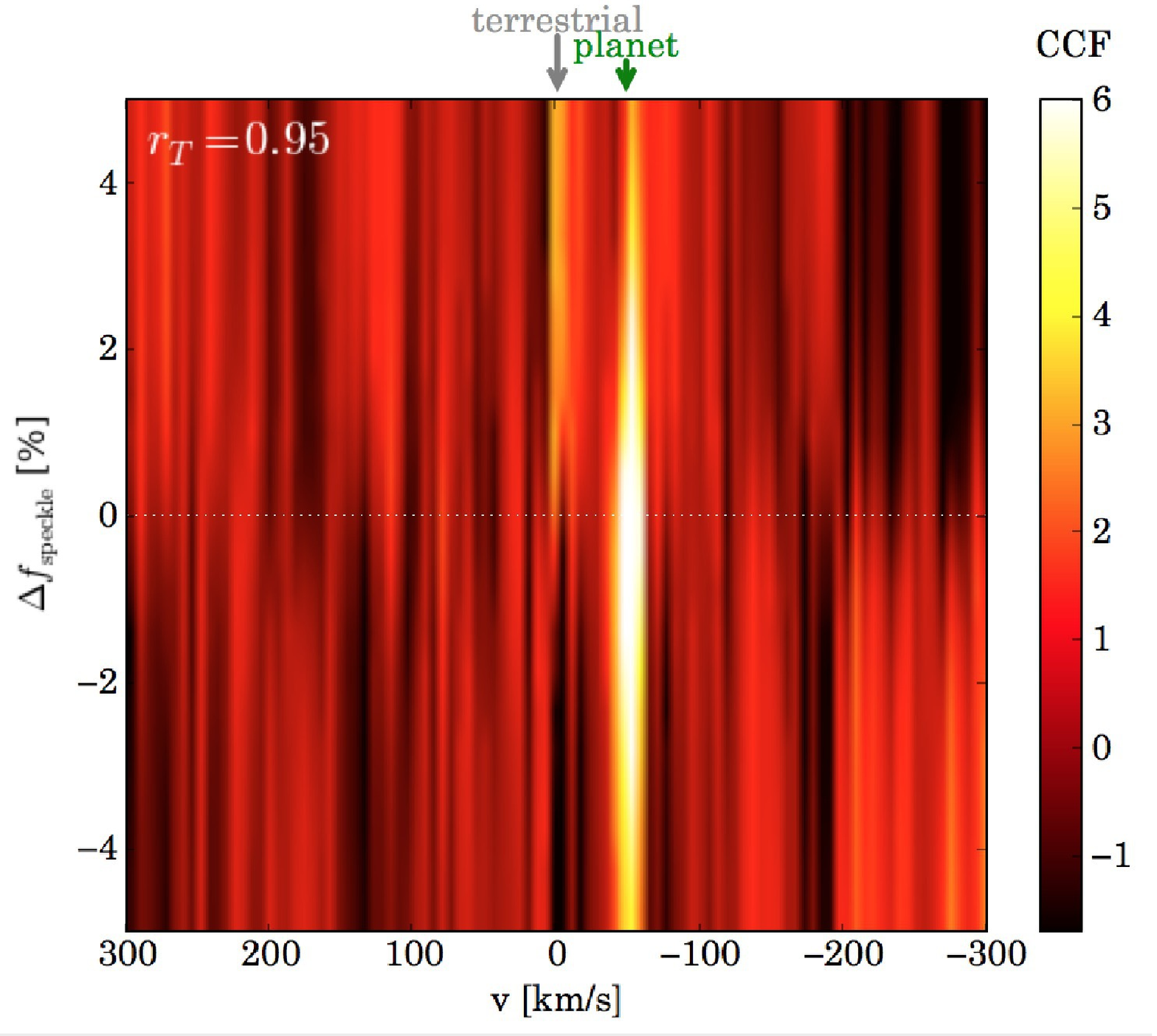}
\caption{CCF for spectra including the systematics of the speckle subtraction and the telluric correction as a function of the deviation of the speckle subtraction, $\Delta f_\mathrm{speckle}$. We assume that the maximum deviation of the telluric line correction is 5 \% ($r_T=0.95$; see text). The planetary system is the same as in Figure \ref{fig:vwater} ($T_\star = 3250 K$).  \label{fig:sys}}
\end{center}
\end{figure}

Figure \ref{fig:sys} shows the CCF determined by assuming these systematic errors to be a function of  $\Delta_\mathrm{speckle}$. As the deviation increases, the signal from telluric water at $v=0$ km/s becomes larger because the telluric absoprtion in the speckle contributes to this signal. This signal can be separated from the planetary one by using the Doppler shift. However, a large deviation $> 3$ \% also distorts the planetary signal. To avoid this fatal effect, the speckle subtraction should be within $\sim 3$ \% for this case, though the required level depends highly on the strength of the signal. We also confirm that the subtraction of the nighttime airglow is not crucial because of its faintness.

\subsection{Oxygen 1.27 $\mu$m and Carbon dioxide 1.6 $\mu$m band}

\begin{figure}
\begin{center}
  \includegraphics[width=\linewidth]{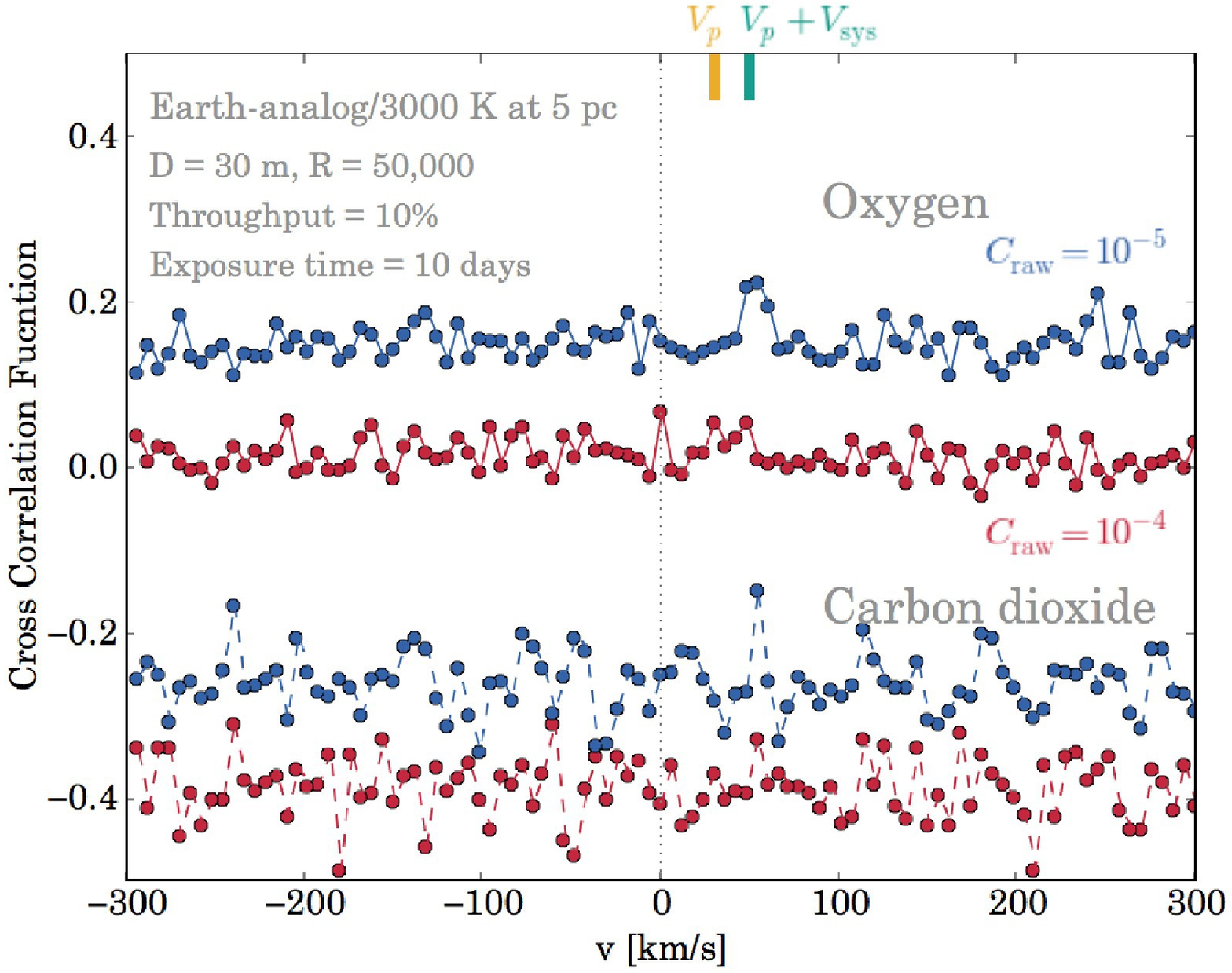}
\caption{CCFs with the oxygen 1.27 $\mu$m (upper two curves) and the carbon dioxide 1.6 $\mu$m (lower two curves) exopnential model. The system is the same as in Figure \ref{fig:vwater} ($T_\star = 3000 K$). The bottom (red) and top (blue) curves correspond to $\Cinst=10^{-4}$ and $\Cinst=10^{-5}$, respectively. The top and bottom curves are artificially shifted by 0.015 and -0.015 in the $y-$ direction for clarity of presentation.\label{fig:ox}}
\end{center}
\end{figure}

The CCFs for the oxygen 1.27 $\mu$m band (denoted by o1.3 in Figure \ref{fig:mixspectra}) are shown in Figure \ref{fig:ox}. We use the exopnential model with $u_c=4 \times 10^{-26}  \mathrm{cm^{-1}/cm^{-2}}$, which approximately reproduces the 1.27 $\mu$m feature of the Earth-analog. Because the oxygen column density is in the AFGL atmospheric constituent profile, the criterion for oxygen is $\sim$ 100 times smaller than that for water detection. The raw contrast $\Cinst$ should be more ambitious for oxygen detection with the CCF analysis. Performing an S/N estimate in the same way as in the water vapor case, we obtain 6 $\sigma$ for $\Cinst = 10^{-5}$, corresponding to $\NCCF \sim 10$. To clearly detect the oxygen 1.27 $\mu$m feature, a raw contrast of $\Cinst=10^{-5}$ is required for this case.

We compare those results with \citet{2012ApJ...758...13K}, in which they considered the detectability of the 1.27 $\mu$ m band in the ELP around a late-type star using low-resolution spectroscopy. They found that if the ExAO + coronagraph + post-processing can directly get an image of the planet, then the photon counts are sufficiently high to detect the 1.27 $\mu$m feature. Because the nighttime airglow is the largest source of background photons and is comparable to the planet flux (see their Figure 7) based on their assumptions, the calibration of the time-variable telluric oxygen emission is crucial. Furthermore, they require an increase in the contrast on the order of $10^2-10^4$ by post-processing. The method presented in this paper requires a long exposure time, however, the photon counts of the nighttime airglow are negligible because we have speckles that are a factor of hundreds to thousands larger than the planet flux. The raw contrast $10^{-5}$ is sufficient for the method with no sophisticated post-processing.  Hence, detection with Doppler-shifted oxygen lines is more robust than that with low-resolution spectroscopy.

The 1.6 $\mu$m band produced by $\mathrm{CO_2}$ (denoted by cd1.6 in Figure \ref{fig:mixspectra}) is the third strongest molecular absorption  of the Earth in this band. We also perfrom a CCF analysis for $\mathrm{CO_2}$ using the exoponential model. We adopt that $u_c = 10^{-23} \mathrm{cm^{-1}/cm^{-2}}$ because the carbon dioxide column density is $\sim$10 times smaller than the water column density. As shown in  Figure \ref{fig:ox}, the $\mathrm{CO_2}$ signal is weaker than those of water and oxygen and is barely detectable (2 $\sigma$ detection) if we assume $\Cinst=10^{-5}$ for the case of $T_\star=3000$ K.

\subsection{Detectability of Scattered Stellar Lines \label{sec:ssl}}

For the extraction of scattered stellar lines, the stellar spectrum itself is used as the template $\ccp(\lambda) = F_\star(\lambda)$. This template naturally includes the velocity of the system, $v=\Vsys$. Hence, one find the CCF signal at $v=\Vp$ because of the relative velocity of the star and the planet.  For the stellar scattered line, the rotational broadening resulting from the stellar spin is crucial. The projected rotational velocity of hundreds of nearby M dwarfs has been determined observationally. \cite{2012AJ....143...93R} and \cite{2003ApJ...583..451M} provided a distribution of $\Vrot \sin i$ for M0-M4.5 and M4-L6, respectively. Whereas most M0-M3 stars (corresponding to $T_\star \gtrsim 3500$ K) have $\Vrot \sin i < $ 10 km/s, M3.5-M6.5 ($T_\star = 2800-3500$ K) stars exhibit a wide range of $\Vrot \sin i$, typically 0-30 km/s. Based on these results, we consider four stellar rotational velocity, $\Vrot=2,5,10,$ and 20 km/s. 

\begin{figure}
\begin{center}
  \includegraphics[width=\linewidth]{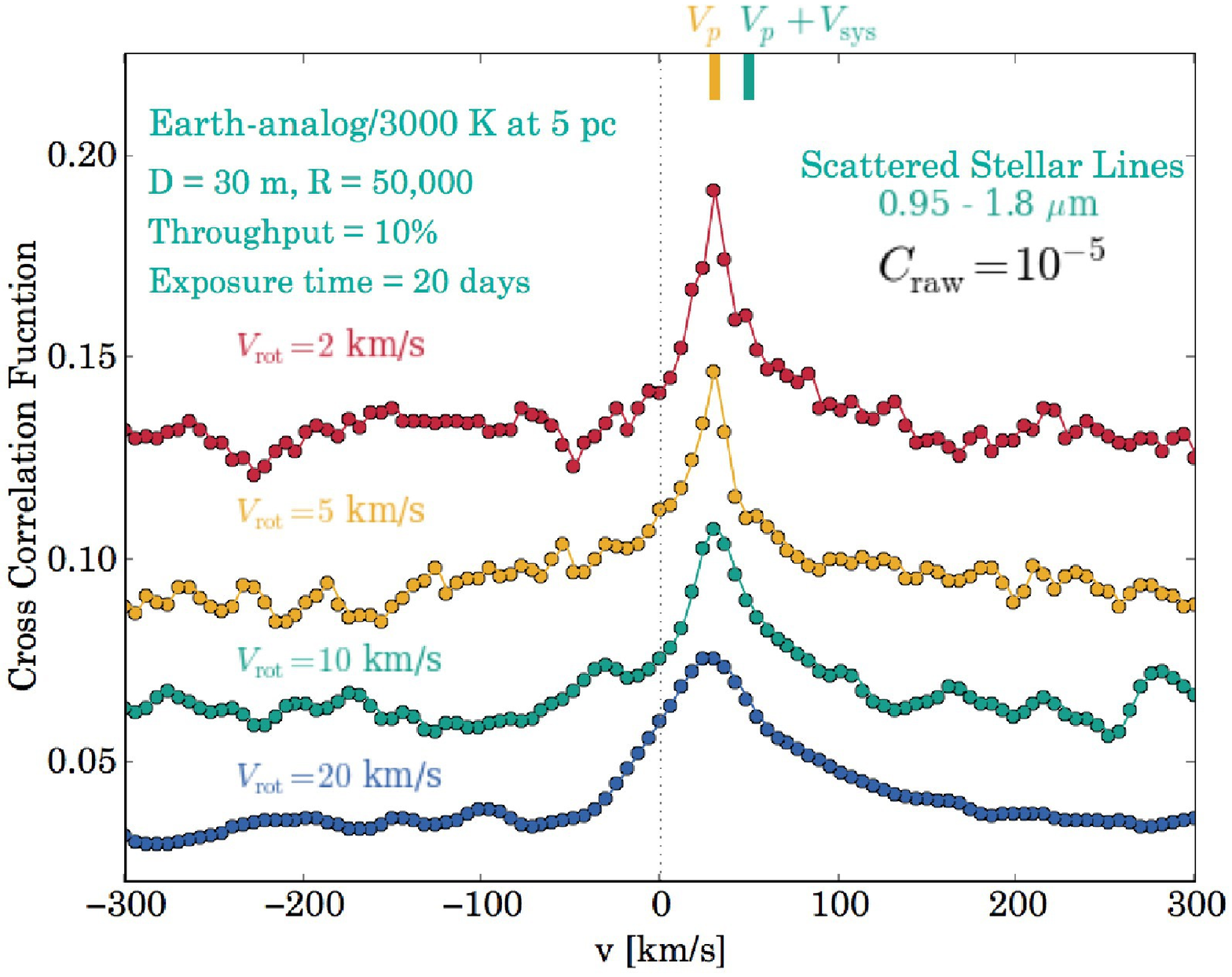}
\caption{CCFs with the stellar spectrum. The red, yellow, green and blue curves correspond to $\Vrot=2,5,10,$ and 20 km/s. We assume $\Cinst=10^{-5}$ and use 0.95-1.8 $\mu$ m. \label{fig:fs}}
\end{center}
\end{figure}

In this simulation, we find that the CCF exhibits a clear feature of the scattered stellar lines at $V = V_p$ as expected (Figure \ref{fig:fs}). Because we assume $R=50,000$, the difference in the width of the CCF signal between $\Vrot=2$ km/s and 5 km/s cannot be resolved. For $\Vrot=$10 or 20 km/s, the signal width roughly corresponds to $\sqrt{2} \times$ 10 or 20 km/s because of the convolution of the widths of the template and the data. This broadening effect weakens the CCF signal. Hence a slow rotator is more suitable as a target.  

Although scattered stellar lines do not have the information on the planetary atmosphere, their detection will be direct evidence of scattered light from the planet. Scattered stellar lines also have the advantage that the signal does not directly depend on the atmospheric composition of the planet. Given the blind search for planets, this virtue is notable. 

\section{Discussion and Summary}

As examined in the previous section, the method needs a long exposure to detect the signal. What are the advantages of this method compared to the low-resolution spectroscopy by direct imaging from the ground ? One is that the method requires no additional post-processing to improve the contrast from the raw contrast by ExAO + coronagraph; in other words, this method is the post-processing itself. This feature is helpful because, for low-resolution spectroscopy of ELPs from the ground, post-processing must be used to improve the final planet-star contrast level to $10^{-8}$ from the raw contrast level.  As far as we know, there has yet to be a full simulation that presents evidence that the ExAO + coronagraph with post processing improves the contrast by more than $10^{-3}-10^{-4}$ from the raw contrast (given $\Csp = 10^{-8}$ and $\Cinst=10^{-4} - 10^{-5}$ used in our simulations). Another point is the simpleness of the calibrations, as discussed in \S 4.2 and 4.3. Low-resolution spectroscopy requires very careful calibrations of the nighttime airglow and the transmittance of our Earth because those are directly related to the estimate of the depth of the absorption feature for the low-resolution case \citep{2012ApJ...758...13K}. Hence, even after post-processing reaches the planet-star contrast of ELPs, the method is complementary and will make molecular detection robust. 

So far we have used a wide range of wavelength 0.95-1.8 $\mu$m. From the standpoint of instrumental development, one might start from a narrower range.  In the previous section, we showed that the J band has the strongest water signal within 0.95-1.8 $\mu$m (Figure \ref{fig:color}). The J band also has the oxygen 1.27 $\mu$m band in it. We also note that the performance of the ExAO will be higher around the J and H bands owing to the wavelength dependence of the wavefront error. In our simulation, the CCFs using the J band exhibited 3 and 12 $\sigma$ water detections for $\Cinst = 10^{-4}$ and $10^{-5}$, respectively. For these reasons, we suggest the J band as the most fruitful band for this method. 

In this paper, we assumed a long exposure (10 days) for characterization of the ELP. Hence, the method might not be suitable for a survey. The point we should consider is how we choose the adequate candidates for a long observing campaign. Direct imaging using ELTs is one promising way to find targets \citep[e.g.][]{2006SPIE.6272E..0NM,2010SPIE.7735E..84M,2010SPIE.7735E..81K,2012SPIE.8447E..1XG,2013A&A...551A..99C,2014arXiv1407.5099M}. Assuming the use of ELTs and the PIAACMC coronagraph, \citet{2012SPIE.8447E..1XG} presented technological solutions to observing rocky planets around nearby M dwarfs. \citet{2013A&A...551A..99C} presented the detectability of planets by assuming an ExAO on an ELT and contrast improvement by post-processing. Utilizing frequencies of low mass planets from the Kepler mission, he found that $\sim$ 10 planets with $ R_p = 1-8 R_\odot $ with radiative equilibrium temperature $T_\mathrm{eq} \le 400$ K can be accessible with ELTs. Those planets will be excellent targets for the method presented here. If the planet is detected by using both radial velocity and astrometry, a long observation at the planet position predicted by those detection methods is possible without direct imaging. However, when only the radial velocity is available, the method becomes more challenging. For this case, we have no information on the position angle. To apply the method to those planets, integral field spectroscopy with $R \sim 50,000$ is required.

The detection of water lines does not mean the presence of surface liquid water. Regarding the search for liquid water, the diagnosis proposed by \citet{2013ApJ...765...76F} is a promising option. They suggested the difference in the diurnal variability of water vapor and oxygen lines as evidence for the existence of surface liquid water. In this paper, we did not include the surface inhomogeneity in our simulations and we might require information on the rotation period to stack the signal to extract the variability. More detailed simulations are required to explore the possibility of further characterization of ELP with the combination of the high-dispersion and high-contrast instruments.

In summary, we investigated a method to characterize both nontransiting and transiting ELPs via the Doppler-shifted water vapor and oxygen lines using high-dispersion spectroscopy and high-contrast instruments on ELTs. This method requires no additional post-processing from the raw contrast. Performing mock observations using the radiative transfer code for the Earth, we examined the feasibility of the method with ELTs. A long observing campaign with a total exposure of 10 days can detect the water vapor lines on nearby ELPs around M-type stars if the high contrast instruments suppress the speckle to the level of $10^{-4} - 10^{-5}$ at 15 mas. If the raw contrast reaches $10^{-5}$, the oxygen 1.27 $\mu$m feature is also detectable. For an ELP around solar-type stars, one need a contrast that is a factor of several hundreds greater at $\sim$ 100 mas. A combination of high-dispersion and high-contrast instruments on ELTs will enable us to characterize nearby exoplanets even for Earth-sized planets within the HZ.

 This work is supported by the Astrobiology Project of the CNSI, NINS (AB261006), and by a Grant-in-Aid for Scientific Research from the JSPS and from the MEXT (No. 25800106). T.H. is supported by Japan Society for Promotion of Science (JSPS) Fellowship for Research (No. 25-3183). We thank Takayuki Kotani, Naoshi Murakami, Taro Matsuo, Nemanja Jovanovic, and Olivier Guyon for fruitful discussions.


\end{document}